\documentclass[]{aa}  

\usepackage{color}
\usepackage{float}
\definecolor{azul}{rgb}{0,0,1}
\usepackage{graphicx}
\usepackage{lscape}
\usepackage{multirow}
\newcommand{\degree}{\ensuremath{^\circ}}
\usepackage{float}
\usepackage{txfonts}
\usepackage{natbib}

\begin{document}

\title{Dust environment and dynamical history of a sample of short period comets}

\titlerunning{Short Period Comets}
\authorrunning{Pozuelos et al.}


   \author{F.J. Pozuelos\inst{1,3}, F. Moreno\inst{1}, F. Aceituno\inst{1}, V. Casanova\inst{1}, A. Sota\inst{1}, J.J. L\'opez-Moreno\inst{1},
          J. Castellano\inst{2}, E. Reina\inst{2}, A. Diepvens\inst{2}, A. Betoret\inst{2}, B. Häusler\inst{2},
          C. González\inst{2}, D. Rodríguez\inst{2}, E. Bryssinck\inst{2},
          E. Cortés\inst{2}, F. García\inst{2}, F. García\inst{2}, F. Limón\inst{2},
          F. Grau\inst{2}, F. Fratev\inst{2}, F. Baldrís\inst{2}, F. A. Rodriguez\inst{2}, F. Montalbán\inst{2}, 
          F. Soldán\inst{2}, G. Muler\inst{2}, I. Almendros\inst{2}, J. Temprano\inst{2}, J. Bel\inst{2},
          J. Sánchez\inst{2}, J. Lopesino\inst{2}, J. Báez\inst{2}, J. F. Hernández\inst{2}, J. L. Martín\inst{2},
          J. M. Ruiz\inst{2}, J. R. Vidal\inst{2}, J. Gaitán\inst{2}, J. L. Salto\inst{2}, J. M. Aymamí\inst{2},
          J. M. Bosch\inst{2}, J. A. Henríquez\inst{2}, J. J. Martín\inst{2}, J. Lacruz\inst{2},
          L. Tremosa\inst{2}, L. Lahuerta\inst{2}, M. Reszelsky\inst{2}, M. Rodríguez\inst{2}, M. Camarasa\inst{2}, M. Campas\inst{2},
          O. Canales\inst{2}, P.J. Dekelver\inst{2}, Q. Moreno\inst{2}, R. Benavides\inst{2}, R. Naves\inst{2},
          R. Dymoc\inst{2}, R. García\inst{2}, S. Lahuerta\inst{2}, T. Climent\inst{2} 
          }

    \institute{Instituto de Astrof\'isica de Andaluc\'ia (CSIC),
              Glorieta de la Astronom\'ia s/n, 18008 Granada, Spain\\
              \email{pozuelos@iaa.es}
         \and
             Amateur Association Cometas-Obs, Spain
        \and
             Universidad de Granada-Phd Program in Physics and Mathematics (FisyMat).\\
             }

   \date{Received March 5, 2014; accepted May 1, 2014}

 
  \abstract
   {}
   {In this work, we present an extended study of the dust environment of a sample of short period
     comets and their dynamical history. With this aim, we characterized the dust tails
     when the comets are active, and we made a statistical study to determine
     their dynamical evolution. The targets selected were 22P/Kopff, 30P/Reinmuth 1, 78P/Gehrels 2, 
     115P/Maury, 118P/Shoemaker-Levy 4, 123P/West-Hartley, 157P/Tritton, 185/Petriew, and 
      P/2011 W2 (Rinner).}
   {We use two different observational data: a set of images taken at the Observatorio de Sierra Nevada and the
    $Af\rho$ curves provided by the amateur astronomical association
  \emph{Cometas-Obs}. To model these
     observations, we use our Monte Carlo dust tail code. From this analysis, we derive the dust parameters, which best describe 
     the dust environment: dust loss rates, ejection velocities, and size distribution of particles.
     On the other hand, we use a numerical integrator to study the dynamical history of the comets, which allows us 
     to determine with a 90\% of confidence level the time spent by these objects in the region of Jupiter Family Comets. 
            }
    {From the Monte Carlo dust tail code, we derived three categories attending to the amount of dust emitted: Weakly active (115P, 157P, and Rinner),
     moderately active (30P, 123P, and 185P), and highly active (22P, 78P, and 118P). The dynamical studies showed that the comets of this sample are young in the 
     Jupiter Family region, where the youngest ones are 22P ($\sim100$ yr), 78P ($\sim500$ yr), and 118P ($\sim600$ yr).
     The study points to a certain correlation between comet activity and time spent in the Jupiter Family region, although this trend is not always fulfilled.
     The largest particle sizes are not tightly constrained, so that the total dust mass derived should be regarded as lower limits.}
   {}

   \keywords{ comets: general --  
              comets: individual: 22P, 30P, 78P, 115P, 118P, 123P, 157P, 185P, and P/2011 W2 --
              methods: observational --
              methods: numerical --
              }

   \maketitle
%

\section{Introduction}

   According to the current theories, comets are the most volatile and 
   least processed materials
   in our Solar System, which was formed from the primitive nebula 4.6 Gyr ago. They
   are considered as fundamental building blocks of giant planets and might 
   be an important source of water on Earth \citep[e.g.][]{hartogh2011}.
   For these reasons, comet research is a hot topic in science today, and quite a 
   few spacecraft missions were devoted to their study. \emph{Giotto} to 1P/Halley \citep{keller1986}, \emph{Deep Space 1} to 19P/Borrelly \citep{soderblom2002}, \emph{Stardust} to
   81P/Wild 2 \citep{brownlee2004}, \emph{Deep Impact} to 9P/Temple 1 \citep{ahearn2005}, and the 
   current \emph{Rosetta} mission on its way to 67P/Churyumov-Gerasimenko
   \citep{schwehm1998} are some examples. It is well known that the importance of the study 
   of short periods comets, which are also
   called Jupiter Family Comets, because they offer the possibility to be 
   studied during several passages near perihelion when the 
   activity increases, which allows us to determine the dust environment and its evolution  
   along the orbital path. This information is necessary to constraint the models describing 
   evolution of different comet families and their contribution to the interplanetary dust
   \citep{sykes2004}.
   In this work, we focus on nine Jupiter Family Comets: 22P/Kopff, 30P/Reinmuth 1, 78P/Gehrels 2,
   115P/Maury, 118P/Shoemaker-Levy 4, 123P/West-Hartley, 157P/Tritton, 185P/Petriew, and P/2011 W2 (Rinner) 
   (hereafter 22P, 30P, 78P, 115P, 118P, 123P, 157P, 185P, and Rinner, respectively). The perihelion distance, 
   aphelion distance, orbital period, and latest perihelion date are displayed in table \ref{table:1}.
   The analysis we have done consists of two different parts: the first one is a dust characterization
   using our Monte Carlo dust tail code, which was developed by \citet{moreno2009} and was used successfully on
   previous studies (e.g. \citet{moreno2012}, from which we adopt the results for comet 22P, see below).
   This procedure allows us to derive the dust parameters: mass loss rates,
   ejection velocities, and size distribution of particles (i.e. maximum size, minimum size, and the power index of the 
   distribution $\delta$). We can also obtain information on the emission pattern, specifically on the emission anisotropy. 
   For the cases where we determine that the emission is 
   anisotropic, we can establish the location of the active areas on the surface and the rotational parameters, as introduced
   by \citet{sekanina1981}, which are the obliquity of the orbit plane to the cometary equator, $I$, and the
   argument of the subsolar meridian at perihelion, $\phi$. The second part in our study is the analysis of the recent (15 Myr)
   dynamical history for each target. To perform this task, we use the numerical integrator developed
   by \citet{chambers1999} as did by other authors before \citep[e.g.,][]{hsieh2012a, hsieh2012b,lacerda2013}.This will serve to
   derive the time spent by each comet in each region and, specifically, in the Jupiter Family region, where 
   it is supposed that the comets become active periodically.
   For some of these comets, this is the first available study to our knowledge.

\begin{table}
\caption{Targets list.}             
\label{table:1}      
\centering                          
\begin{tabular}{c c c c c}        
\hline\hline 
\noalign{\smallskip}                
\multirow{2}{*}{Comet} & q & Q & Period & Last perihelion \\
  & (AU) & (AU) & (yr) & date \\    
\hline 
\noalign{\smallskip}
   22P & 1.57 & 5.33 & 6.43 & May 25, 2009 \\                  
   30P & 1.88 & 5.66 & 7.34 & April 19, 2010 \\      
   78P & 2.00 & 5.46  & 7.22 & January 12, 2012 \\
   115P & 2.03 & 6.46 & 8.76 & October 6, 2011\\
   118P & 1.98 & 4.94 & 6.45 & January 2, 2010\\
   123P & 2.12 & 5.59 & 7.59  & July 4, 2011\\ 
   157P & 1.35 & 5.46 & 6.31 &  February 20, 2010\\      
   185P & 0.93 & 5.26 & 5.46 & August 13, 2012 \\
   Rinner & 2.30 & 5.29 & 7.40 & November 6, 2011 \\
\hline                                   
\end{tabular}
\end{table}


\section{Observations and data reduction}

   \begin{figure*}
   \centering
   \includegraphics[width=0.6\textwidth]{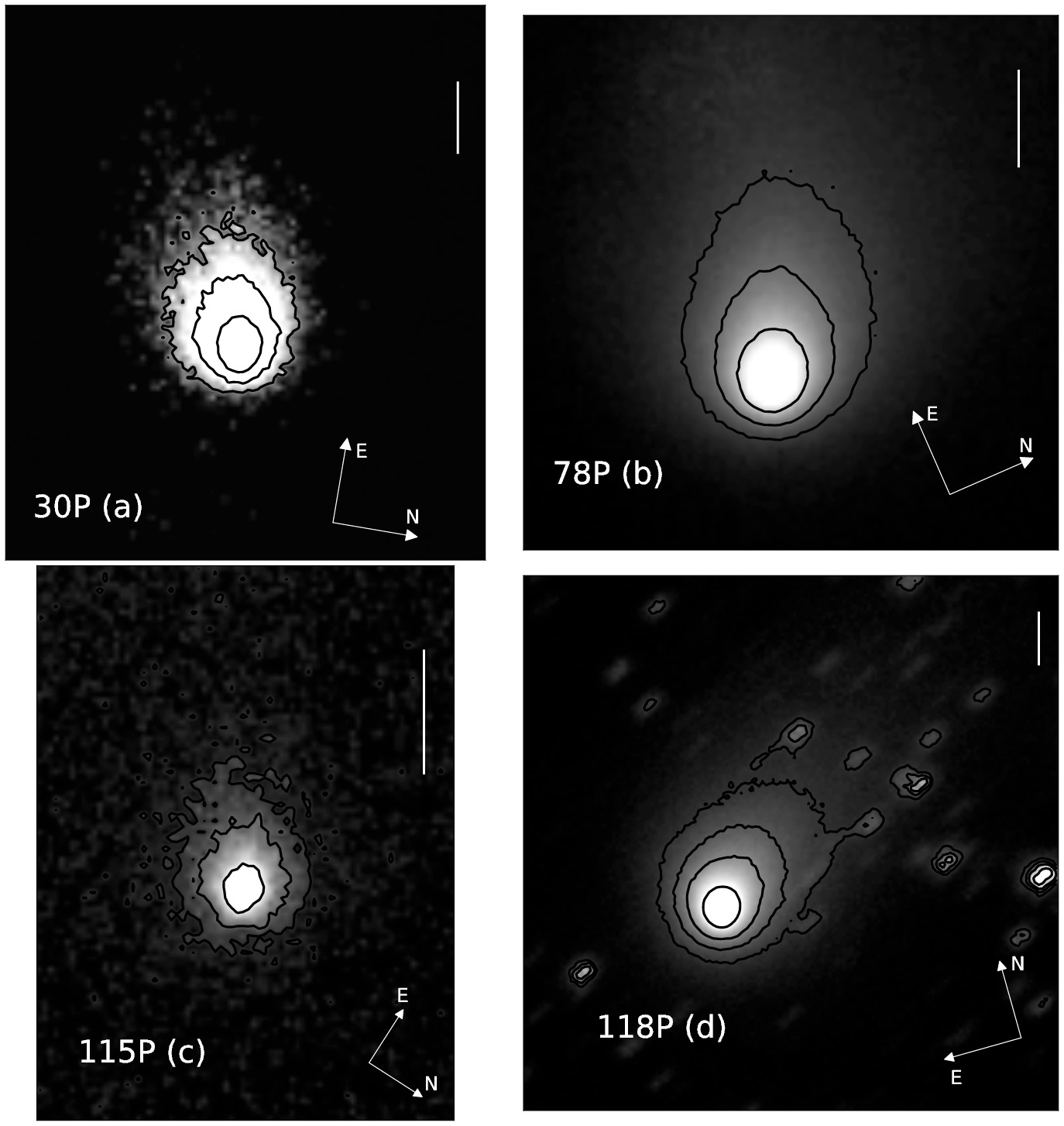}
   \caption{Observations obtained using a CCD camera at 1.52 m telescope of the Observatorio
            de Sierra Nevada in Granada, Spain. (a) 30P/Reinmuth 1 on May 15, 2010. Isophote levels in Solar Disk Units (SDU) are 
             2.00$\times{10^{-13}}$, 0.75$\times{10^{-13}}$, and 0.25$\times{10^{-13}}$. (b) 78P/Gehrels 2 on December 19, 2011.
             Isophote levels are 0.55$\times{10^{-12}}$, 2.65$\times{10^{-13}}$, and 1.35$\times{10^{-13}}$. (c) 115P/Maury on July 15, 2011.
             Isophote levels are 1.00$\times{10^{-13}}$, 3.00$\times{10^{-14}}$, and 1.30$\times{10^{-14}}$. (d) 118P/Shoemaker-Levy 4 on 
             December 12, 2009. Isophote levels are 1.50$\times{10^{-13}}$, 6.00$\times{10^{-14}}$, 3.50$\times{10^{-14}}$, and 2.00$\times{10^{-14}}$.
             In all cases, the directions of celestial North and East are given. The vertical bars correspond to 10$^{4}$km in the sky.}
              \label{real1}%
    \end{figure*}
    
     \begin{figure*}
   \centering
   \includegraphics[width=0.6\textwidth]{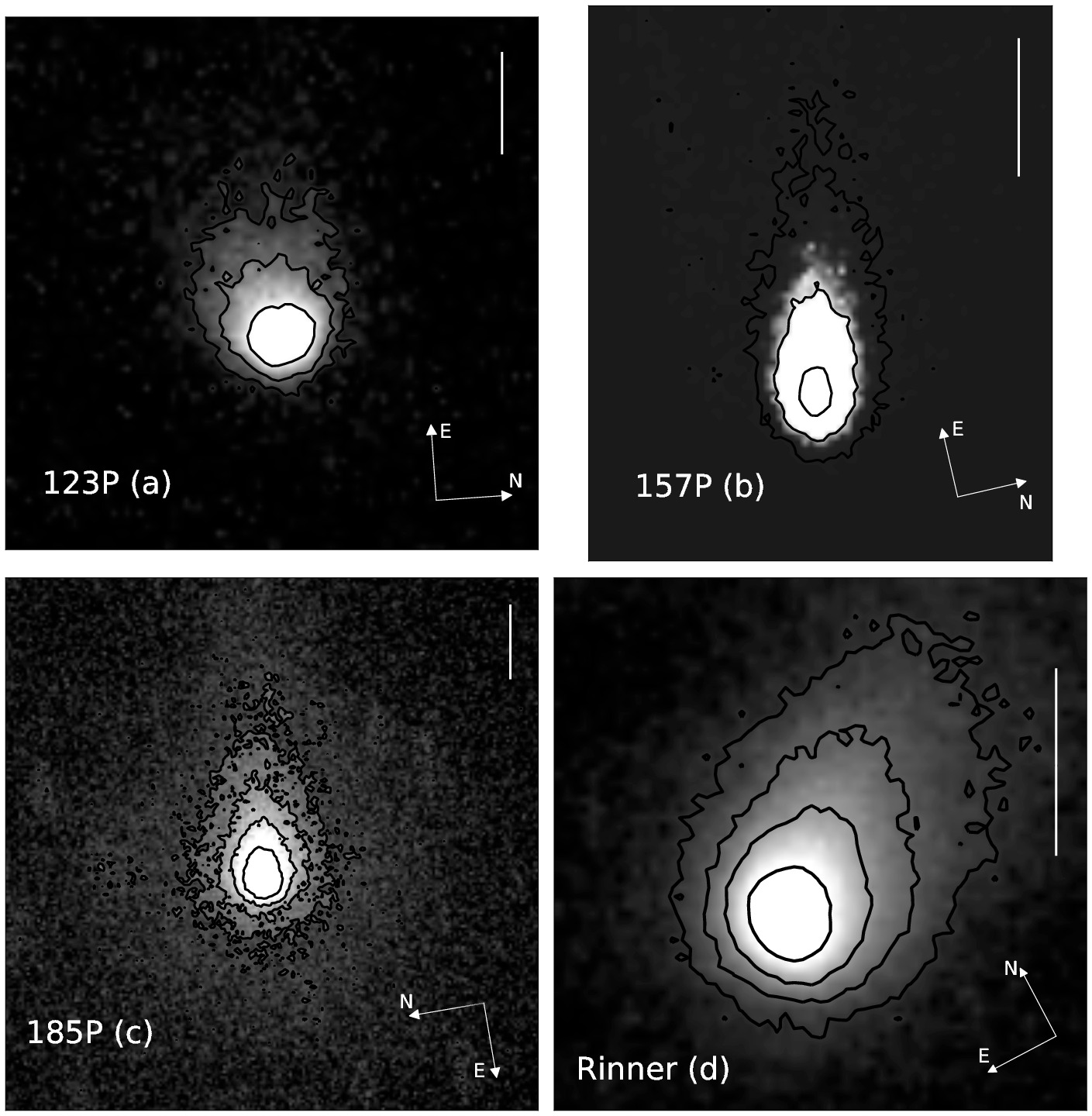}
   \caption{Observations obtained using a CCD camera at 1.52 m telescope of the Observatorio
            de Sierra Nevada in Granada, Spain. (a) 123P/West-Hartley on February 26, 2011. Isophote levels in Solar Disk Units (SDU) are 
             1.00$\times{10^{-13}}$, 0.35$\times{10^{-13}}$, and 0.15$\times{10^{-13}}$. (b) 157P/Tritton on March 10, 2010.
             Isophote levels are 6.00$\times{10^{-13}}$, 0.75$\times{10^{-13}}$, and 2.65$\times{10^{-14}}$. (c) 185P/Petriew on July 15, 2012.
             Isophote levels are 1.80$\times{10^{-13}}$, 1.00$\times{10^{-13}}$, 0.60$\times{10^{-13}}$, and 0.35$\times{10^{-13}}$. (d) P/2011 W2 (Rinner) on 
             January 4, 2012. Isophote levels are 6.00$\times{10^{-14}}$, 2.70$\times{10^{-14}}$, 1.50$\times{10^{-14}}$, and 0.80$\times{10^{-14}}$.
             In all cases, the directions of celestial North and East are given. The vertical bars correspond to 10$^{4}$km in the sky.}
              \label{real2}%
    \end{figure*}
   The first block of our observation data were taken at the 1.52 m telescope of Sierra Nevada 
   Observatory (OSN) in Granada, Spain.                                    
   We used a 1024$\times$1024 pixel CCD camera with a Johnson red filter to minimize gaseous emissions. 
   The pixel size in the sky was 0."46, so the field of view was 7'.8$\times$7'.8.
   To improve the signal-to-noise 
   ratio, the comets were imaged several times using integration times in the range 60-300 sec. 
   The individual images at each night were bias subtracted and 
   flat-fielded using standard techniques. The flux calibration was made using the USNO-B1.0 
   star catalog \citep{monet2003}.The individual images of the comets were calibrated to mag arcsec$^{-2}$
   and then converted to solar disk intensity units (hereafter SDU). After calibration, 
   the images corresponding to each single night were shifted to a reference image by taking their
   apparent sky motion into account, and 
   then a median of those images was taken. For the modeling
   purposes, the final images are rotated to the photographic plane $(N,M)$ 
   \citep{finson&probstein1968a}, where the Sun is toward $-M$.  
   Table \ref{table:2} shows the log of the observations. Negative values in time to perihelion correspond to pre-perihelion observations.
   Representative images are displayed in Figs. \ref{real1} and \ref{real2}.
 
   The second block of observational data correspond to the $Af\rho$ curves around perihelion date ($\sim{300}$ days). 
   These observations were
   carried out by the amateur astronomical association \emph{Cometas-Obs}. The $Af\rho$ measurements are 
   presented as a function of the heliocentric distance and are always referred to an
   aperture of radius $\rho=10^4$ km projected on the sky at each observation 
   date. The calibration of the \emph{Cometas-Obs} observations was performed
   with the star catalogs CMC-14 and USNO A2.0.
   
\begin{table*}
\caption{Log of the OSN observations.}             
\label{table:2}      
\centering          
\begin{tabular}{c c c c c c c c}     
\hline\hline       
\noalign{\smallskip}                     
\multirow{2}{*}{Comet} & Observation Date & $r_{h}$ $^{1}$ & $\Delta$ & Resolution & Phase & Position &$Af\rho$ ($\rho=10^{4}km$) $^{2}$  \\
& (UT) & (AU) & (AU) & (km pixel$^{-1}$)& Angle ($\degree$) &Angle ($\degree$) & (cm) \\  
\hline  
\noalign{\smallskip}                  
\multirow{2}{*}{30P/Reinmuth 1} & 2010 Mar 10 21:45& -1.916 & 1.579 & 526.8 & 31.1 & 87.1 & 52 \\ \
& 2010 May 15 21:10 & 1.898 & 2.147 & 716.3 & 28.0 & 99.6 & 61 \\ \hline
\noalign{\smallskip}
\multirow{2}{*}{78P/Gehrels 2} & 2011 Dic 19 20:00 & -2.018 & 1.647 & 549.5 & 28.9 & 66.5 &380 \\ \
& 2012 Jan 4 20:15 & -2.009 & 1.805 & 602.2 &  29.2 & 67.0 & 470 \\ \hline
\noalign{\smallskip}
115P/Maury & 2011 Jul 2 22:00 & -2.146 & 1.343 & 448.0 & 18.5 & 122.7 & 17\\ \hline
\noalign{\smallskip}
118P/Shoemaker-Levy 4 & 2009 Dec 12 01:45 & -1.991 & 1.032 & 1377.2 & 8.9 & 324.9 & 103\\ \hline
\noalign{\smallskip}
\multirow{2}{*}{123P/West-Hartley} & 2011 Feb 26 23:00 & -2.346 & 1.970 & 657.3 & 24.5 &86.4 & 40 \\ \
& 2011 Mar 31 21:00 & 2.253 & 2.252 & 751.3 & 25.6 &85.4 & 50 \\ \hline
\noalign{\smallskip}
157P/Tritton & 2010 Mar 10 21:30 & 1.376 & 1.343 & 448.0 & 42.8 & 77.0 & 20\\ \hline
\noalign{\smallskip}
185P/Petriew & 2012 Jul 15 03:15 & -1.027 & 1.097 & 366.0 & 57.0 & 260.2 & 17\\ \hline
\noalign{\smallskip}
\multirow{2}{*}{P/2011 W2 (Rinner)} & 2011 Dic 22 03:00 & 2.326 & 1.451 & 484.1& 14.0 & 309.9 & 18 \\ \
& 2012 Jan 4 02:00 & 2.340 & 1.412 & 471.2 & 10.2 & 332.2 & 22 \\ \hline     
\end{tabular}
\tablefoot{ \\
            $^{1}$ Negative values correspond with pre-perihelion, positive values with post-perihelion.\\
            $^{2}$ The $Af\rho$ values for phase angle $\leq30\degree$ have been corrected according to the equation (2) (see text).}\\  
            
\end{table*}


\section{Monte Carlo dust tail model}

   The dust tail analysis was performed by the Monte Carlo dust tail code, which allows us to fit the OSN images
   and the observational $Af\rho$ curves provided by \emph{Cometas-Obs}.
   This code has been successfully used on previous works on characterization of 
   dust environments of comets and Main-belt comets, such as 29P/Schwassmann-Wachmann 1
   and P/2010 R2 (La Sagra) \citep{moreno2009,moreno2011}. This code is 
   also called the Granada model \citep[see][]{fulle2010} in the dust studies for the Rosetta mission 
   target, 67P/Churyumov-Gerasimenko. The model describes the motion of the particles when 
   they leave the nucleus and are submitted to the gravity force of the Sun and the radiation
   pressure, so that the trajectory of the particles around the Sun is Keplerian. The $\beta$ parameter is defined as the ratio
   of the radiation pressure force to the gravity force and is given for spherical particles as 
   $\beta=C_{pr}Q_{pr}/(\rho_{d}d)$, where $C_{pr}=1.19\times10^{-3}$ km m$^{-2}$; $Q_{pr}$ 
   is the scattering efficiency for radiation pressure, which is $Q_{pr}\sim1$ for large absorbing
   grains \citep{burns1979}; $\rho_{d}$ is the mass
   density, assumed at $\rho_{d}=10^{3}$ kg m$^{-3}$; and $d$ is the particle diameter. 
   We use Mie theory for the interaction of the electromagnetic field with the spherical 
   particles to compute the geometric albedo, $p_{v}$, and $Q_{pr}$. The parameter, $p_{v}$, is a
   function of the phase angle $\alpha$ and the particle 
   radius. We assume the particles as glassy carbon spheres of refractive index $m=1.88+0.71i$ \citep{edoh1983} 
   at $\lambda=0.6$ $\mu$m. We compute a
   large number of dust particle trayectories and calculate their positions on the $(N,M)$ plane and 
   their contribution to the tail brightness. The free parameters of the model are
   dust mass loss rate, the ejection velocities of the particles, the size distribution, and 
   the dust ejection pattern.

   \subsection{$Af\rho$}

\begin{figure}
      \includegraphics[width=1\columnwidth]{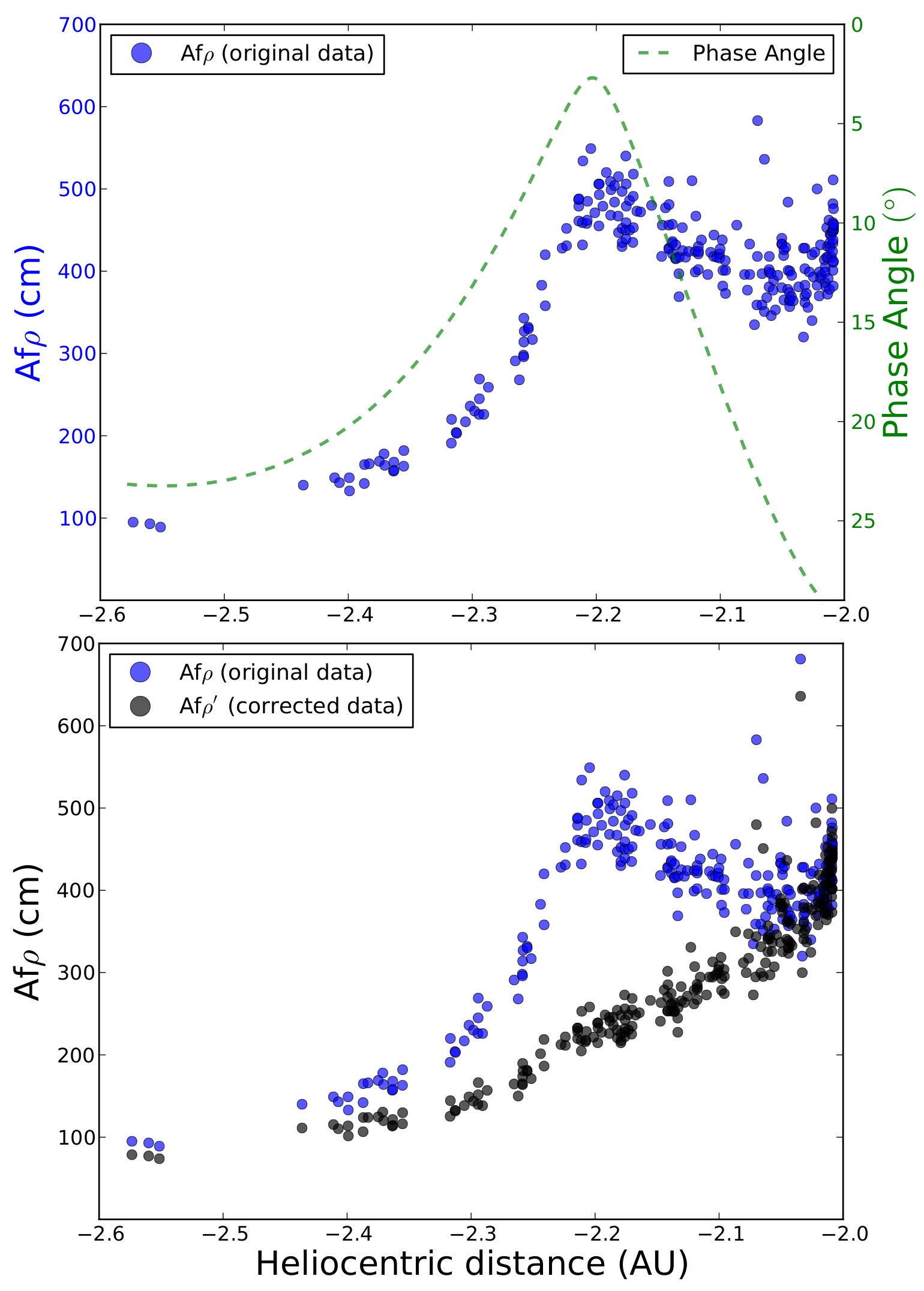}
          \caption{$Af\rho$ pre-perihelion measurements of comet 78P/Gehrels 2 provided by \emph{Cometas-Obs}. Upper panel:
                  original $Af\rho$ measurements and phase angle as a function of 
                  heliocentric distance. Lower panel: $Af\rho$ and $Af\rho^{\prime}$ after 
                  the backscattering effect correction using equation (2) as function of heliocentric distance.}
           \label{sto}
   \end{figure}

   The $Af\rho$ quantity [cm] \citep{ahearn1984} is related to the dust coma brightness, 
   where $A$ is the dust geometric albedo, $f$ the filling factor 
   in the aperture field of view (proportional to the dust optical thickness), 
   and $\rho$ is the linear radius of aperture at the
   comet, which is the sky-plane radius. When the cometary coma is in steady-state,
   $Af\rho$ is independent of the observation coma 
   radius $\rho$ if the surface brightness of the dust coma is proportional to $\rho^{-1}$. 
   It is formulated as follows:
\begin{equation}
Af\rho = {4r_{h}^{2}\Delta^{2} \over \rho} \frac{F_c}{F_s}\;,
\end{equation}
  where $r_{h}$ is the heliocentric distance and $\Delta$ the geocentric 
  distance. The parameter, $F_c$, is the measured cometary flux integrated within a
  radius of aperture $\rho$, and $F_s$ is the total solar flux.
  For each comet, we have the $Af\rho$ curve as heliocentric distance function
  provided by \emph{Cometas-Obs} for an aperture of radius $\rho=10^{4}$ km, $\sim300$ days around perihelion, and the $Af\rho$ measurements derived from the OSN 
  observations with the same aperture. Some of the $Af\rho$ data correspond to times, where the phase angle was close to  zero degree, so that the backscattering 
  enhancement became apparent \citep{kolokolova2004}. We could not model this enhancement: for the assumed absorbing spherical particles, the phase function 
  is approximately constant except for the forward spike for particles whose radius is r$\geq\lambda$. Then, we corrected these enlarged $Af\rho$ at small phase angles
  by assuming a certain linear phase coefficient, which we apply to the data at phase angles $\alpha\leq30\degree$. We adopted a linear phase coefficient  of
  0.03 mag $deg^{-1}$, which is in the range 0.02-0.04 mag $deg^{-1}$ estimated by \citet{meech1987} from various comets. In this way, the corrected $Af\rho^{\prime}$ values
  are computed as a function of the original $Af\rho$ values at phase angle $\alpha$ as:
\begin{equation}
Af\rho^{\prime} = 10^\frac{-\beta(30-\alpha)}{2.5} Af\rho \; .
\end{equation}
 To illustrate this correction, we show in its application to comet 78P/Gehrels 2 Fig. \ref{sto}. In the upper panel, the correlation of the original $Af\rho$ data
 with the phase angle and the lower panel the final $Af\rho$ curve after correcting those values by equation (2) is seen. The same equation is applied to the OSN images when the phase angle
 is $\alpha\leq30\degree$ (see table \ref{table:2}).

\section{Dust analysis}

As described in the previous section, we use our Monte Carlo dust tail code to retrieve the dust properties of each
comet in our sample. The code has many important parameters, so that a number of simplifying assumptions should be made to make
the problem tractable. The dust particles are assumed spherical with a density of 1000 kg m$^{-3}$ and a refractive index of 
$m=1.88+0.71i$, which is typical of carbonaceous spheres at red wavelengths \citep{edoh1983}. This gives a geometric albedo of $p_{v}=0.04$
for particle sizes of r$\gtrsim1 \mu m$ at a wide range of phase angles. The particle ejection velocity is parametrized as $v(t,\beta)=v_{1}(t)\times\beta^{1/2}$,
where $v_{1}(t)$ is a time dependent function to be determined in the modeling procedure. In addition, the emission pattern, which are possible spatial asymmetries
in the particle ejection, might appear. The asymmetric ejection pattern is parametrized by considering a rotating nucleus with active areas on it, whose rotating
axis is defined by the obliquity, $I$, and the argument of the subsolar meridian at perihelion, as defined in \citet{sekanina1981}. The rotation 
period, $P$, is not generally constrained if the ejecta age is much longer than $P$, which is normally the case. The particles are assumed distributed
broadly in size, so that the minimum size is always set in principle in the sub-micrometer range, while the maximum size is set in the centimetre range. The size distribution is
assumed to be given
by a power law, $n(r)\propto r^{-\delta}$, where $\delta$ is set to vary in the -4.2 to -3 domain, which is the range that has been determined for other comets \citep[e.g.,][]{jockers1997}.
All of those parameters, $v_{1}(t)$, $r_{min}$, $r_{max}$, and $\delta$, and the mass loss rate are a function of the heliocentric distance, so that some kind of dependence on $r_{h}$
must be established. In addition, the activity onset time should also be specified. On the other hand, current knowledge of physical 
properties of cometary nuclei established the bulk density below $\rho=1000$ kg m$^{-3}$ \citep{carry2012}. Values of $\rho=600$ kg m$^{-3}$ have been reported 
for comets 81P/Wild 2 \citep{davidsson2004} and Temple 1 \citep{ahearn2005}, so we adopted that value. The ejection velocity at a distance R$\sim 20 R_{N}$, 
where R$_{N}$ is the nucleus radii and R is the distance where the gas drag vanishes, should overcome the escape velocity, which is given by $v_{esc}=\sqrt{2GM/R}$. Assuming a spherically-shaped nucleus, we get 
$v_{esc}=R_{N}\sqrt{(2/15)\pi \rho G}$, where $\rho=600$ kg m$^{-3}$. In cases where $R_{N}$ has been estimated by other authors, the minimum ejection velocity should
verify the condition  $v_{min}\gtrsim v_{esc}$. Considering that the minimum particle velocity determined in the model is 
$v_{min}\sim v_{esc}$, we can give an upper limit estimate of the nucleus radius in all the other cases.

The Monte Carlo dust tail code, is a forward code whose output is a dust tail image corresponding to a given set of input parameters.
Given the large amount of parameters, the solution is likely not unique: approximately the same tail brightness can be likely
achieved by assuming another set of input parameters. However, if the number of available images and/or $Af\rho$ measurements cover a significant 
orbital  arc, it is clear that the indetermination is reduced. Our general procedure first consists in assuming the most simple case: isotropic particle
ejection, $r_{min}=1\mu$m, $r_{max}=1$cm, $v_{1}(t)$ monotonically increasing toward perihelion, $\delta=-3.5$, and $dM/dt$ set to a value, which reproduce the measured tail intensity
in the optocenter, assuming a monotonic decrease with heliocentric distance. From this starting point, we then start to vary the parameters, assuming a certain different
dependence with heliocentric distance, until an acceptable agreement with both the dust tail images and the $Af\rho$ measurements is reached. Then, if
we find no way to fit the data using an isotropic ejection model after many trial-and-error procedures, we switch to the anisotropic model where the active
area location and rotational parameters must be set. 

Using the procedure described above for each comet in the sample, we present the results on the dust parameters organized in the following
way: in tabular form, where the main properties derived of the dust environment of each comet is given (tables \ref{DPI} and \ref{DPII}), and a series of plots 
on the dependence on the heliocentric distance of the 
dust mass loss rate, the ejection velocities for $r=1$ cm particles, the maximum particle size, 
and the power index of the size distribution. Representative plot is shown in the case of the comet 30P in Fig. \ref{30P-DP} and in appendix B, the results for each comet
are individually displayed (see figures \ref{78P-DP} to \ref{rinner-DP}).
In addition, the representative plot of the comparison between observations and model in the case of the comet 30P is shown in Fig. \ref{c30P}, and
the comparison between observations and models for each comet individually are displayed in the 
appendix C (figures \ref{c78P} to \ref{crinner}).

\begin{figure}[]
      \includegraphics[width=1\columnwidth]{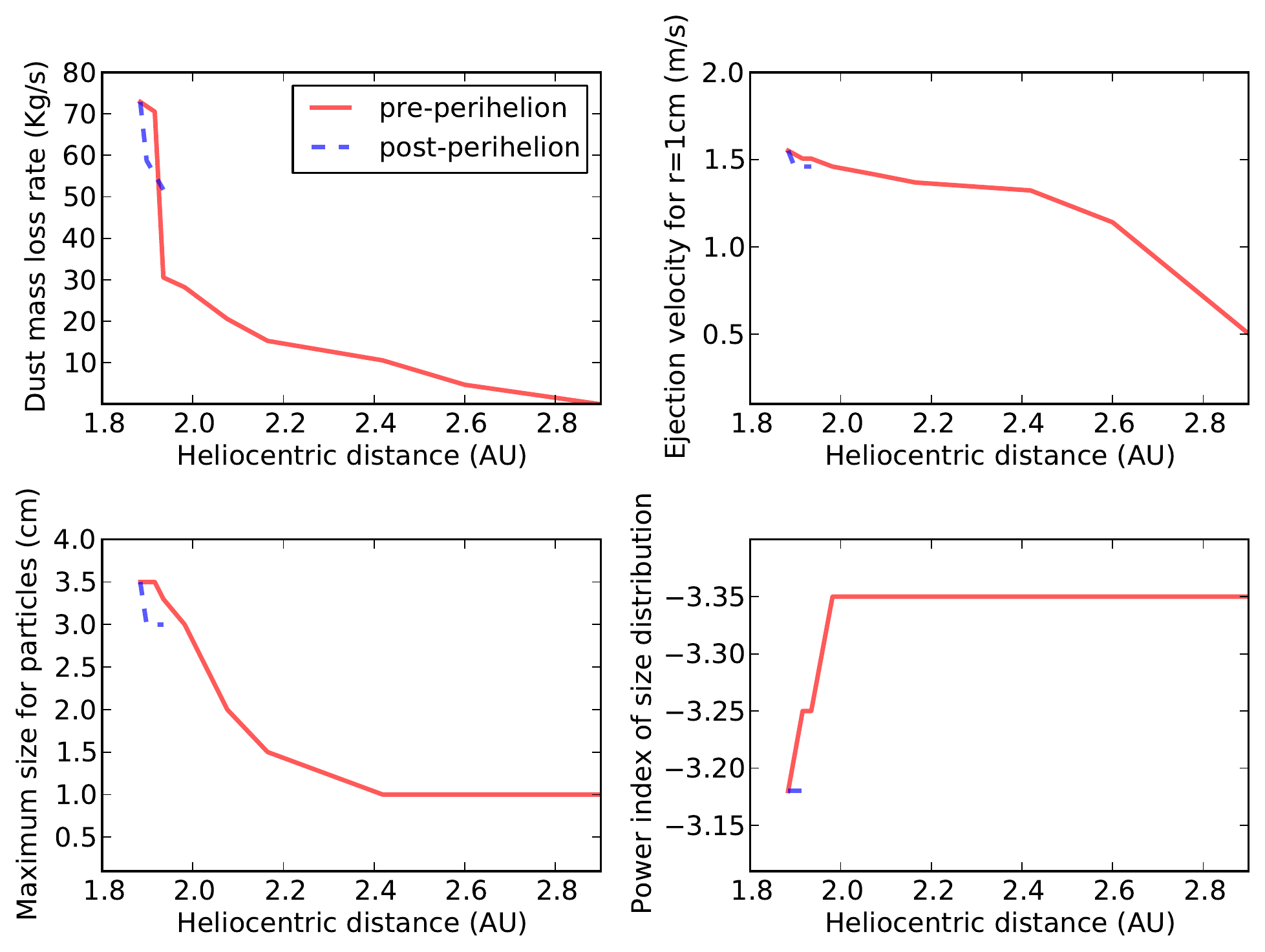}
          \caption{Representative figure of the evolution of the dust parameter evolution obtained in the model versus 
               the heliocentric distance for comet 30P/Reinmuth 1. The panels are as follows:
               (a) dust mass loss rate [kg/s]; (b) ejection velocities
               for particles of r=1 cm glassy carbon spheres [m/s]; (c) maximum size of 
               the particles [cm]; and (d) power index of the size distribution. In all 
               cases, the solid red lines correspond to pre-perihelion and the dashed blue lines
               to post-perihelion. In appendix B, the results for each comet individually are displayed
               (see figures \ref{78P-DP} to \ref{rinner-DP}).
              }
         \label{30P-DP}
   \end{figure}

\begin{figure}[]
   \includegraphics[width=1\columnwidth]{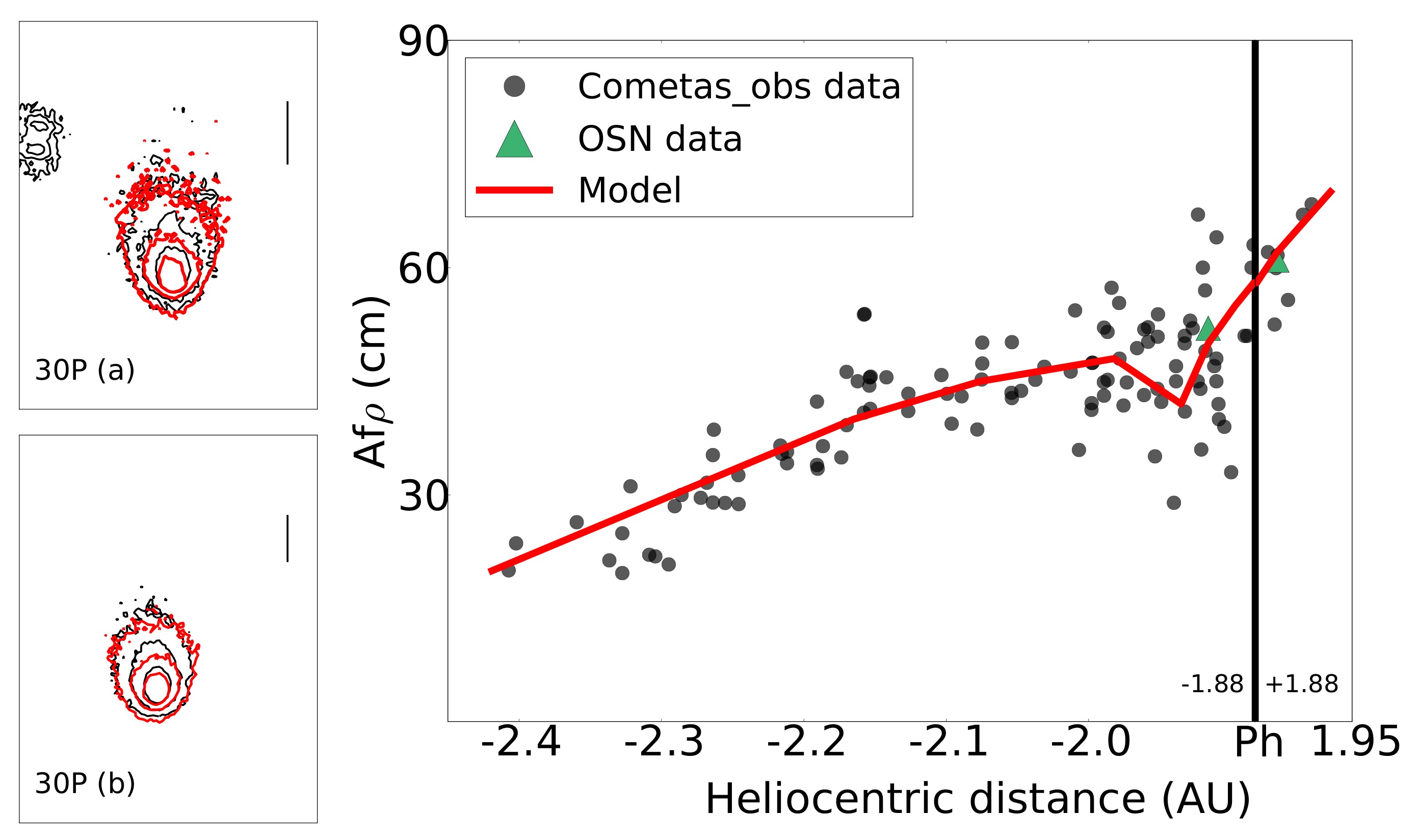}
      \caption{Representative figure of the comparison between observations and model of comet 30P/Reinmuth 1. 
              Left panels: isophote fields (a) March 10, 2010, and (b) May 15, 2010.
              In both cases, isophote levels are 2.00$\times{10^{-13}}$, 0.75$\times{10^{-13}}$, and 0.25$\times{10^{-13}}$ SDU.
              The black contours correspond to the 
              OSN observations and the red contours to the model. Vertical bars 
              correspond to $10^{4}$ km on the sky. Right panel: parameter $Af\rho$ versus heliocentric distance. The black dots correspond to Cometas-Obs 
              data, and the green triangles are the OSN data, which correspond to March 10 and May 15, 2010. The red line is the model. The observations and 
              the model refer to $\rho=10^{4}$km.
              In appendix C, the comparison between observations and models for each comet individually are displayed
              (see figures \ref{c78P} to \ref{crinner}).
              }
         \label{c30P}
   \end{figure}

\subsection{Discussion}

The dust environment of the 22P was already reported by \citet{moreno2012}, where the authors concluded that this comet shows a clear time dependent
asymmetric ejection behavior with an enhanced activity at heliocentric distances beyond 2.5 AU pre-perihelion. This is also accompanied by enhanced particle ejection velocity.
The maximum size for the particles were estimated as 1.4 cm with a constant power index of -3.1. The peak of dust mass loss rate and the peak of ejection velocities
were reached at perihelion with values $Q_{d}=260$ kg s$^{-1}$ and $v=2.7$ m s$^{-1}$ for 1-cm grains. The total dust lost per orbit was $8\times10^{9}$ kg.
The annual dust loss rate is $T_{d}=1.24\times10^{9}$ kg yr$^{-1}$
, and the averaged dust mass loss rate per orbit is 40 kg/s. The contribution to the interplanetary dust of this comet corresponds to about 0.4$\%$ of the $\sim2.9\times10^{11}$ kg yr$^{-1}$ that must 
be replenished if the cloud of interplanetary dust is in steady state \citep{grun1985}. 

For 30P, 115P, and 157P, we derived an anisotropic ejection pattern with active areas on the nucleus surface (see table \ref{DPI}). In the case of 30P, the rotational parameters, $I$ and $\phi$,
have been taken from \citet{krolikowska1998} as $I=107\degree$ and $\phi=321\degree$. However, for 115P and 157P, these parameters have been derived from the model. The 78P is the most active comet
in our sample with a peak dust loss rate at perihelion with a value of $Q_{d}=530$ kg s$^{-1}$ and a total dust mass ejected of $5.8\times10^9$ kg. This comet was study by \citet{mazzotta2011} in 
its previous perihelion passage on October 2004. The authors estimated that the dust production rate at perihelion with values between $Q_{d}=14-345$ kg s$^{-1}$ using a method derived from the 
one used by \citet{jewitt2009} to compute the dust production rate of active Centarus. They also obtained $Af\rho=846\pm55$ cm in an aperture of radius $\rho=7.3\times10^3$ km, and they concluded
that this comet is more active than the the average Jupiter Family Comets at a given heliocentric distance. In addition, \citet{lowry2003} reported a stellar appearance of 78P at $r_{h}=5.46$ AU pre-perihelion,
and any possible coma contribution to the observed flux was likely to be small or non existent, which is consistent with our model where the comet is not active at such large pre-perihelion distances. 
From our studies, we can classify our targets in three different categories: weakly active comets (115P, 157P, and Rinner) with an average annual dust production rate of $T_{d}<1\times10^{8}$ kg yr$^{-1}$;
moderately active comets (30P, 123P, and 185P) with $T_{d}=1-3\times10^{8}$ kg yr$^{-1}$; and highly active comets (22P, 78P, and 118P) with $T_{d}>8\times10^{8}$ kg yr$^{-1}$. 
It is necessary to consider that we do not have observations after perihelion for the comet 115P and 123P. That is, our observational information covers less than half of the orbit, losing the part of the branch which is supposed
to be the most active. For this reason, our results for these comets are lower limits in the $T_{d}$ measurements.   

\begin{landscape}
\begin{table}
\caption{\label{DPI} Dust properties summary of the targets under study I.}
\begin{tabular}{l c c c c c c c }
\hline\hline
\noalign{\smallskip}
\multirow{2}{*}{Comet}& Emission  & Active areas  & Size distribution & Size distribution & Maximum nucleus & Obliquity & Argument of subsolar\\
 & pattern $^{1}$ &location ($\degree$) & $r_{min},r_{max}$ (cm)& $\delta_{min},\delta_{max}$ &radius (km)& ($\degree$) & meridan at perihelion ($\degree$) \\
\noalign{\smallskip}
\hline
\hline
\noalign{\smallskip}
30P/Reinmuth 1    & Ani (50\%) & -30 to +30 & 10$^{-4}$, 3.5 & -3.35, -3.18 & 3.9 $^{2}$ & 107 $^{3}$ & 133 $^{3}$ \\
78P/Gehrels 2  & Iso (100\%) & -- & 10$^{-4}$, 3.0 & -3.40, -3.05  & 3.6 & -- & --\\
115P/Maury  & Ani (70\%) & -20 to +60 & 10$^{-4}$, 4.0 & -3.13, -3.05 & 4.0 & 25 & 280 \\
118P/Shoemaker-Levy 4 & Iso (100\%) & --     &  10$^{-4}$, 3.0  & -3.20, -3.05 & 2.4 $^{4}$ & -- & --\\
123P/West-Hartley & Iso (100\%) & -- & 10$^{-4}$, 2.5 & -3.32, -3.15 & 2.0 $^{5}$ & -- & -- \\
157P/Tritton & Ani (70\%) & -30 to +30 & 10$^{-4}$, 3.0 & -3.35, -3.15 & 1.6 &  10 & 150 \\
185P/Petriew & Iso (100\%) & -- & 10$^{-4}$, 6.0   & -3.60, -3.00 & 5.7 & -- & --  \\
P/2011 W2 (Rinner) & Iso (100\%) & -- & 10$^{-4}$, 2.5   & -3.20, -3.15 & 2.2 & -- & --  \\ \hline
\end{tabular}
\tablefoot{ \\
            $^{1}$ Iso=Isotropic ejection; Ani=Anisotropic ejection.\\
            $^{2}$ \citet{scotti1994}.\\ 
            $^{3}$ \citet{krolikowska1998}.\\
            $^{4}$ \citet{lamy2004}.\\
            $^{5}$ \citet{tancredi2006}.\\ 
             }
\end{table}

\begin{table}
\caption{\label{DPII} Dust properties summary of the targets under study II.}
 \begin{tabular}{l c c c c c c }
\hline\hline
\noalign{\smallskip}
\multirow{2}{*}{Comet} & Peak dust loss &  Peak ejection velocity &Total dust mass & Total dust mass & Averaged dust & Contribution to the  \\
 & rate (kg/s) &of 1-cm grains (m/s) & ejected (kg) & ejected per year (kg/yr)&mass loss rate (kg/s) & interplanetary dust (\%) $^{1}$ \\
 \noalign{\smallskip}
\hline
\hline
\noalign{\smallskip}
30P/Reinmuth 1    & 73.0 & 1.4 & $8.2\times10^{8}$ & $2.1\times10^{8}$ & 6.8 & 0.07 \\
78P/Gehrels 2  & 530.0 & 2.1 & $5.8\times10^{9}$ & $1.5\times10^{9}$ & 47.5 & 0.52\\
115P/Maury    & 45.0 & 1.4 & $2.9\times10^{8}$ & $6.9\times10^{7}$ & 2.1 & 0.02\\
118P/Shoemaker-Levy 4         &  180.0 & 2.3  & $2.3\times10^{9}$  & $6.5\times10^{8}$  & 20.8 & 0.22\\
123P/West-Hartley & 65.0 & 1.8 & $5.4\times10^{8}$ & $1.4\times10^{8}$ & 4.5 & 0.05 \\
157P/Tritton & 50.0 & 2.1 & $2.9\times10^{8}$ & $8.7\times10^{7}$ & 2.7 & 0.03 \\
185P/Petriew & 143.0 & 7.2  & $9.0\times10^{8}$ & $3.0\times10^{8}$ & 9.6 & 0.10 \\
P/2011 W2 (Rinner) & 15.0 & 1.7  & $2.6\times10^{8}$ & $6.4\times10^{7}$ & 2.0 & 0.02 \\ \hline
\end{tabular}
\tablefoot{ \\
            $^{1}$ Annual contribution to the interplanetary dust replacement \citep{grun1985}.\\
            }
\end{table}
\end{landscape}

\section{Dynamical history analysis}
\citet{levison1994} were the first to make a comprehensive set of long-term integrations (up to $10^{7}$ yr) to study the dynamical evolution
of short period comets. The authors argue that it is necessary to make a statistical study
using several orbits for each comet with slightly different initial orbital elements due to the chaotic nature of each individual orbit. For this reason, the authors made a 10 Myr backward (and forward) integration
for the 160 short period comets know at that time and 3 clones for each comet with offsets in the positions along the $x$, $y$, and $z$ directions of +0.01 AU.
That is, they used 640 test particles for their integrations. They conclude that the long-term integrations into the 
the past or future are statistically equivalent, and they obtained that $\sim92\%$ of the total particles were ejected from the 
Solar System, and $\sim6\%$ were destroyed by becoming Sun-grazers. The median lifetime of Jupiter Family Comets (hereafter JFCs) was derived as $3.25\times10^{5}$ yr. 
In a later study of the same authors \citep{levison1997}, they estimated that the physical lifetime of JFCs is between (3-25)$\times 10^{3}$ yr,
where the most likely value is 12$\times 10^{3}$ yr. 
In our case, we use the Mercury package
version 6.2, a numerical integrator developed by \citet{chambers1999}, to determine the dynamical evolution of our targets, that has been used by other 
authors with the same purpose \citep[e.g.,][]{hsieh2012a,hsieh2012b,lacerda2013}.
Due to the chaotic nature of the targets, which was mentioned in the \citet{levison1994} study, we generate a total of 99 clones having 
2$\sigma$ dispersion in three of the orbital elements: semimajor axis, eccentricity, and inclination (hereafter $a$, $e$, and $i$),
where $\sigma$ is the uncertainty in the corresponding parameter as given in the JPL Horizons on-line Solar System data (see ssd.pjl.nasa.gov/?horizons).
In table \ref{table:or}, we show the orbital parameters and the $1\sigma$ uncertainty 
of our targets extracted from that web page.
These 99 clones plus the real object make 
a total of 100 massless test particles to perform a statistical study for each comet, which supposes 900 massless test particles.
The Sun and the eight planets are
considered as massive bodies.
We used the hybrid algorithm, which combines a second-order mixed-variable
symplectic algorithm with a Burlisch-Stoer integrator to control close encounters.
The initial time step is 8 days, and the clones are removed any time during the integration when they are beyond 1000 AU from the 
Sun.
The total integration time was 15 Myr, which is time enough to determine the most visited regions for each comet
and derive the time spent in the region of JFCs, region that is supposed to be the location where 
the comets reach a temperature high enough to be active periodically. We divide the possible locations of the comets in four regions
attending to their dynamical properties at each moment in the study: JFCs-type with $a<a_{S}/(1+e)$; Centaur-type, confined by $a_{S}/(1+e)<a<a_{N}$ 
and $e<0.8$; Halley-type, which is similar to Centaur-type but with $e>0.8$; and Transneptunian-type with $a>a_{N}$, where $a_{S}$ and $a_{N}$ are
the semimajor axes of Saturn and Neptune, respectively.

In this study, we neglect the non-gravitational forces using the same arguments as \citet{lacerda2013}. 
Thus, assuming that the non-gravitational acceleration, $T$, is due to a single sublimation 
jet tangential to the comet orbit, the change rate of the semimajor axis is described by 

\begin{equation}
da/dt=2Va^{2}T/GM_{\odot}\;,
\end{equation}
where $T$ is the acceleration due to the single jet and is given by
\begin{equation}
T=(dM_{d}/dt)(v_{d}/m_{nuc})\;. 
\end{equation}

In these equations, $V$ is the orbital velocity, $a$ is the semimajor axis, $G$ is the gravitational constant, $M_{\odot}$
is the Sun mass, $v_{d}$ is the dust velocity, and $m_{nuc}$ is the mass of the nucleus.
To show a general justification to neglect the gravitational forces that are valid to our complete list of targets, 
we compute the maximum rate of change of the semimajor axis that corresponds to the comet using the maximum $a$, maximum $dM/dt$, 
maximum $v_{d}$, and minimum $m_{nuc}$. From our comet sample, these values are $a_{max}=4.25$ AU (115P), $(dM/dt)_{max}=47.5$ kg s$^{-1}$ (78P),
 $(v_{d})_{max}=708$ m s$^{-1}$ (185P). The minimum comet nucleus was inferred for 157P as $R_{N}\leq1.6$ km, so we adopt $(R_{N})_{min}=1.6$ km,
which is a minimum nucleus of $(m_{nuc})_{min}=1.03\times 10^{13}$ kg.
Taken all those values together, we get $T=2.2\times10^{-5}$ AU yr$^{-2}$ and $da/dt=4.8\times10^{-5}$ AU yr$^{-1}$. On the other
hand, the lifetime of sublimation from a single jet would be $t_{sub}=6871$ yr, which is based on the nucleus size and $(dM/dt)_{max}$.
Then, the total deviation in semimajor axis would be $(da/dt)_{max}\times t_{sub}=0.33$ AU. This deviation, which should be 
considered as an upper limit, is completely negligible in the scale of variations we are dealing with in the dynamical 
analysis of the orbital evolution. 
This result is close to the one derived for \citet{lacerda2013}, where the author gives the 
maximum semimajor deviation for P/2010 T020 LINEAR-Grauer as 0.42 AU.

As a result of our 15 Myr backward integration for all targets, we find that the $\sim98\%$ of the particles are ejected before the end of the integration, and in almost all cases, the surviving clones
are in the transneptunian region. Thus, we focused on the first 1 Myr of backward in the orbital evolution, where the $\sim20\%$ of the test particles still remain in the Solar System.
This time is enough to obtain a general view 
of the visited regions by each comet. After that, we display the last 5000 yr with a 100 yr temporal resolution
to obtain the time spent by these comets in the JFCs region with a confidence level of $90\%$.

\begin{figure}
      \includegraphics[width=1\columnwidth]{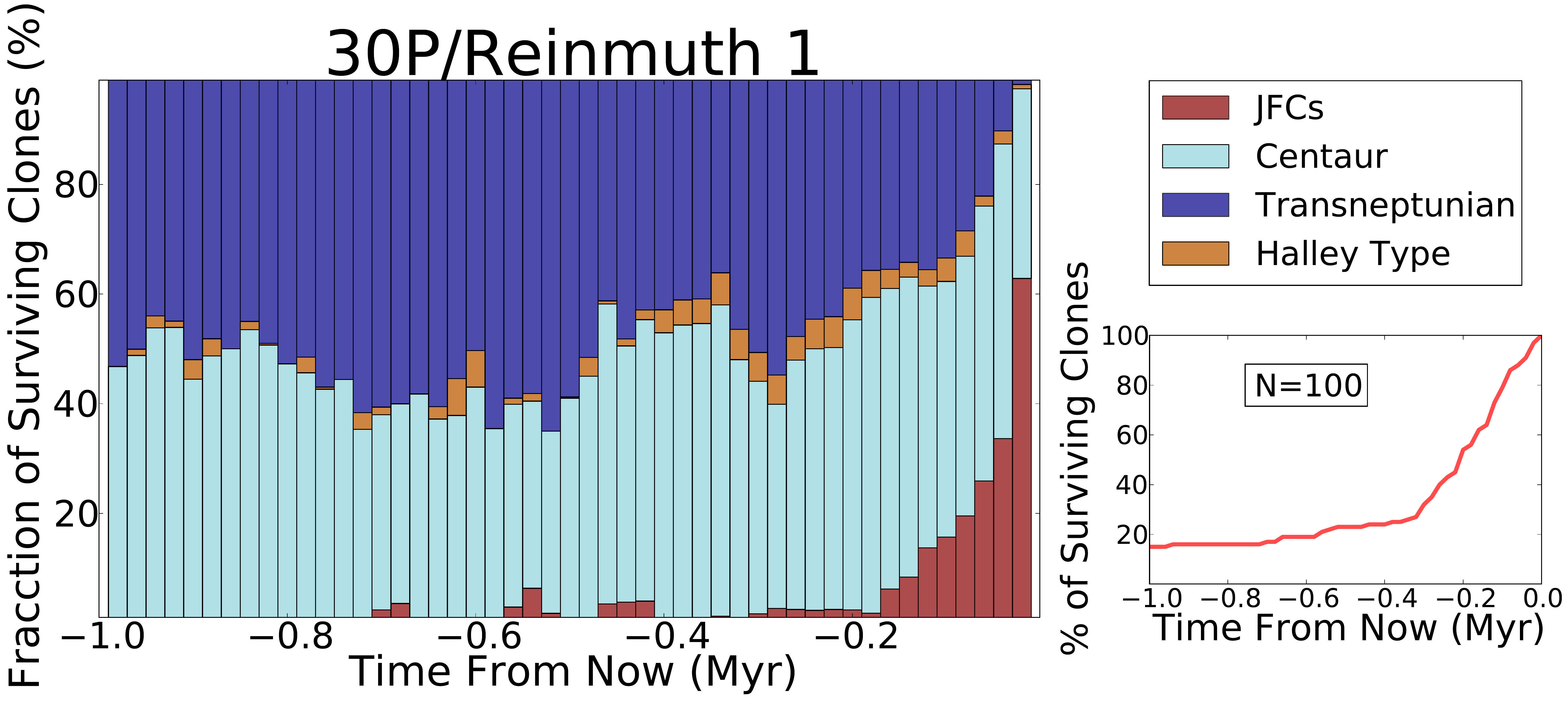}
          \caption{30P/Reinmuth 1 backward in time orbital evolution during 1 Myr. Left panel: fraction of surviving clones ($\%$) versus
          time from now (Myr). The colors represent the regions visited by the
          test particles (red: Jupiter Family region; cyan: Centaur; blue: Transneptunian; yellow: Halley Type). The resolution is $2\times10^{4}$ yr.
          Right bottom panel: the $\%$ of surviving clones versus time from now (Myr), where N=100 is the number of the initial test massless particles.}
         \label{30P-dinamica}
   \end{figure}

\begin{figure}
      \includegraphics[width=1\columnwidth]{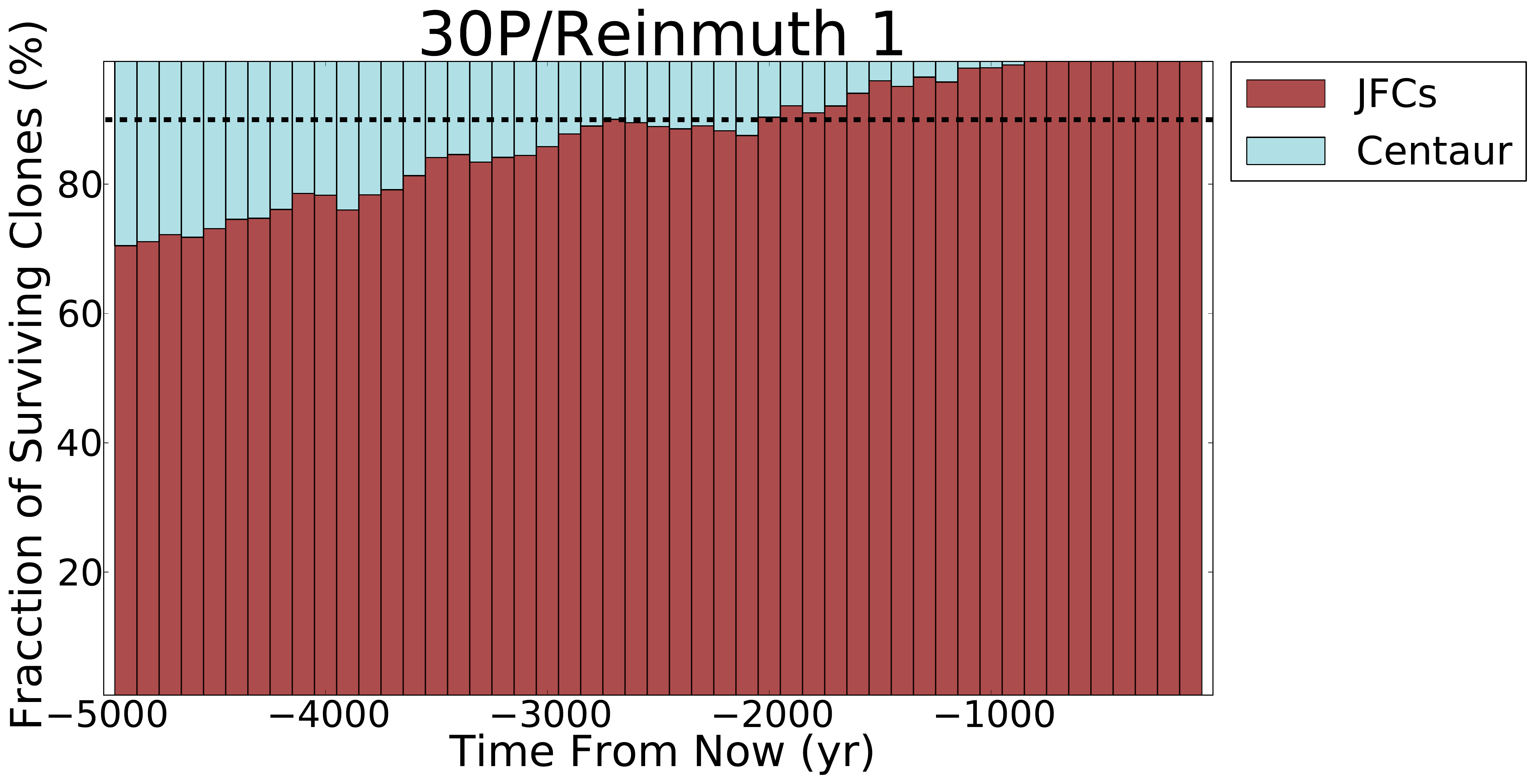}
          \caption{30P/Reinmuth 1 during the last $5\times10^{3} yr$. Fraction of surviving clones ($\%$) versus
          time from now (Myr). The colors represent the regions visited by the
          test particles (red: Jupiter Family region; cyan: Centaur). The dashed line 
          marks the bars with a confidence level equal or larger than 90$\%$ of 
          the clones in the Jupiter Family region. The resolution is 100 yr, and the number of the initial test particles is N=100.
                    }
         \label{30P-CL90}
   \end{figure}

As an example of this procedure, we show the results for 30P (see Fig. \ref{30P-dinamica}) in detail. For this comet, we determined that $85\%$ of the particles were ejected from the 
Solar System after 1 Myr of backward integration. We can see that 
most of the particles stay in the JFCs region during the first $\sim2\times10^{4}$ yr , but they moved on into further regions as centaurs and transneptunian objects at 
$\sim2\times10^{5}$ yr. To determine the time spent by 30P in the JFCs region, we 
show the last $5\times10^{3}$ yr with a resolution of 100 yr in Fig. \ref{30P-CL90}. We derive with 90$\%$ of confidence level that 30P comet spent $\sim2\times10^{3}$ yr in
this location. 

\begin{figure}
      \includegraphics[width=1\columnwidth]{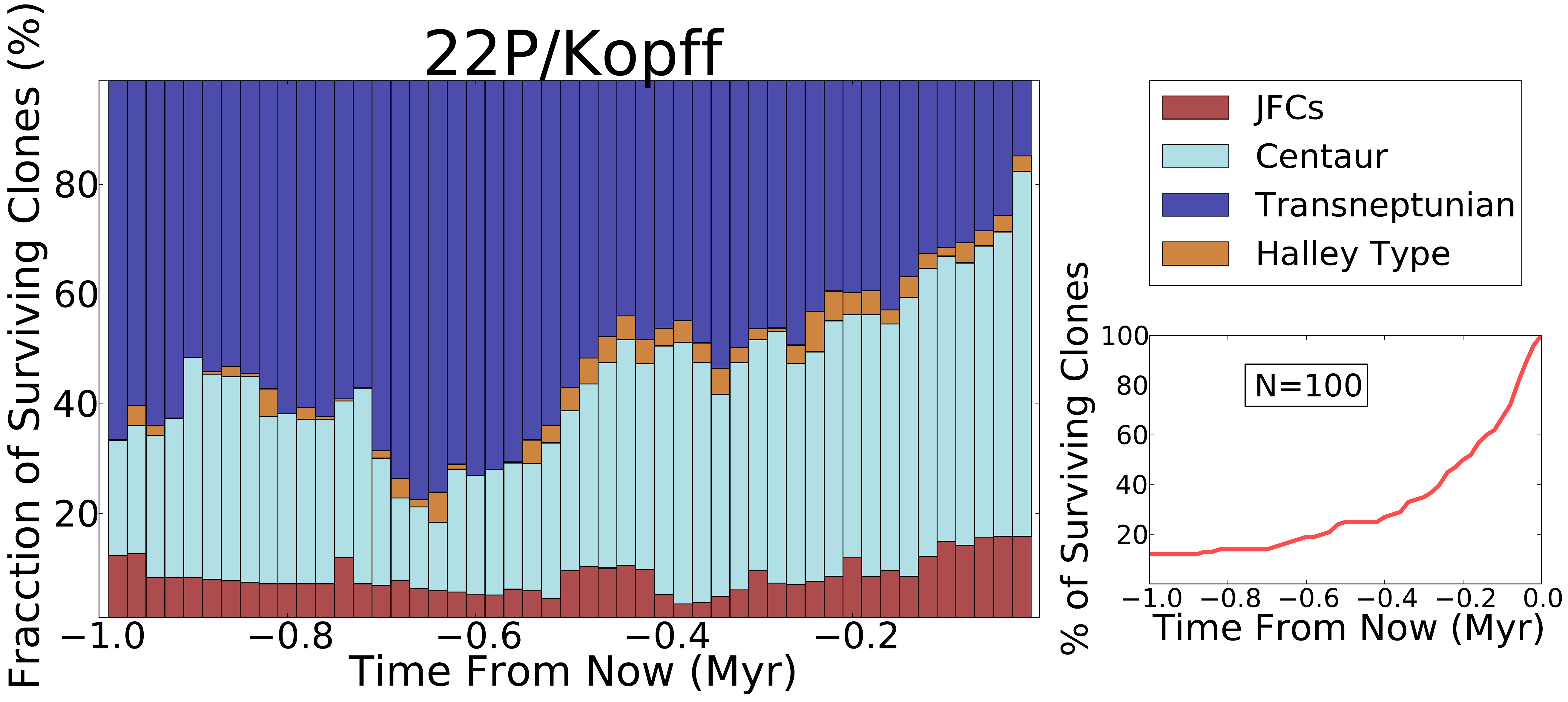}
          \caption{As in Fig. \ref{30P-dinamica}, but for comet 22P/Kopff.
                  }
         \label{22P-dinamica}
   \end{figure}

\begin{figure}
      \includegraphics[width=1\columnwidth]{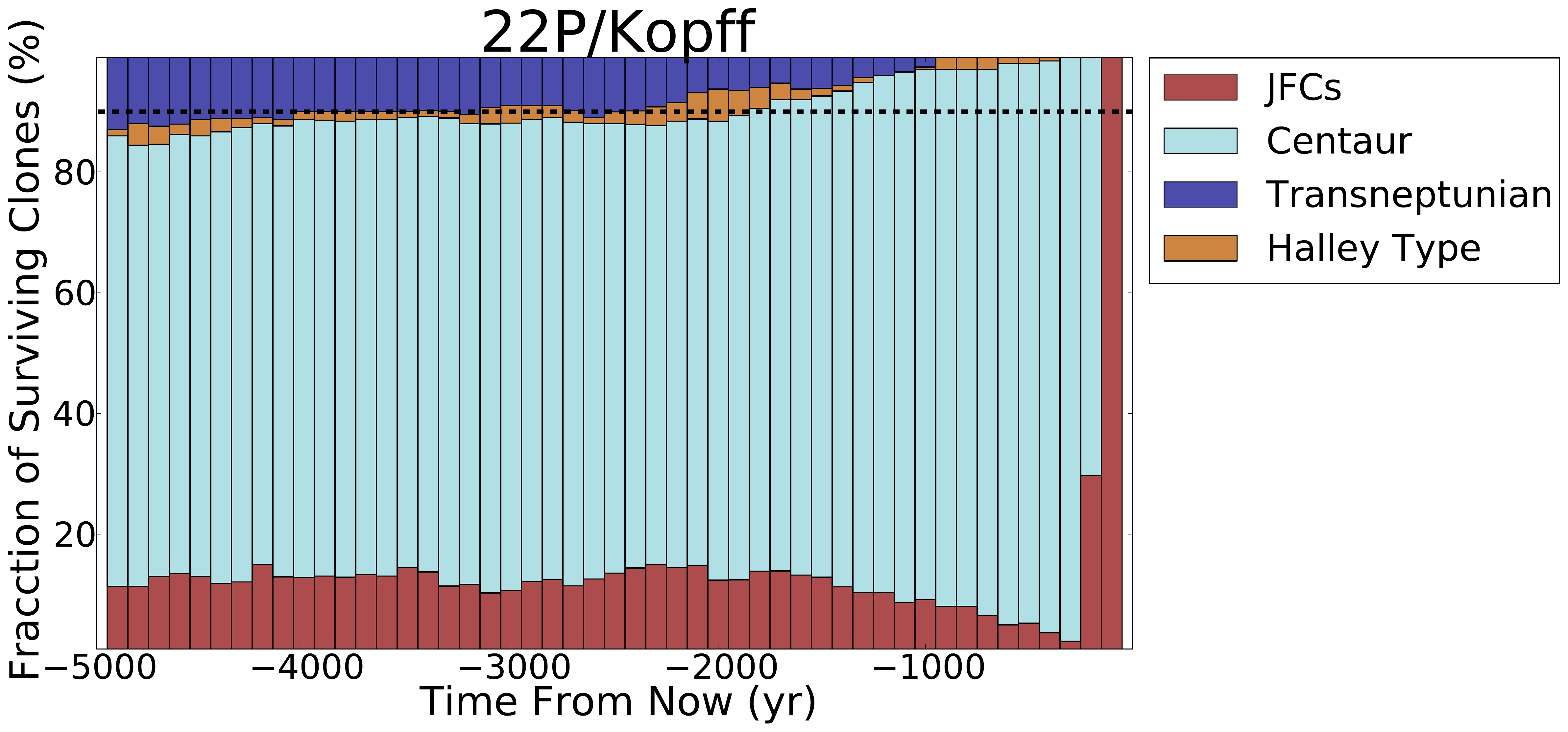}
          \caption{As in Fig. \ref{30P-CL90}, but for comet 22P/Kopff.
                    }
         \label{22P-CL90}
   \end{figure}

A special case within the sample is comet 22P, which turned to be the youngest one in our study. Its dynamical
analysis shows that the $88\%$ of the test particles are ejected from the Solar System 
before 1 Myr. The probability 
to be at the JFCs region in this period remains always under 20$\%$ (Fig. \ref{22P-dinamica}). If we focused on the last $5\times10^{3}$ yr 
(see Fig. \ref{22P-CL90}), we determine the time spent in JFCs region as $\sim100$ yr.
This agrees with its discovery in 1906. It seems that this comet came from the 
Centaur region, which is the most likely region occupied by the object along this period.

\subsection{Discussion}
\begin{figure}
      \includegraphics[width=1\columnwidth]{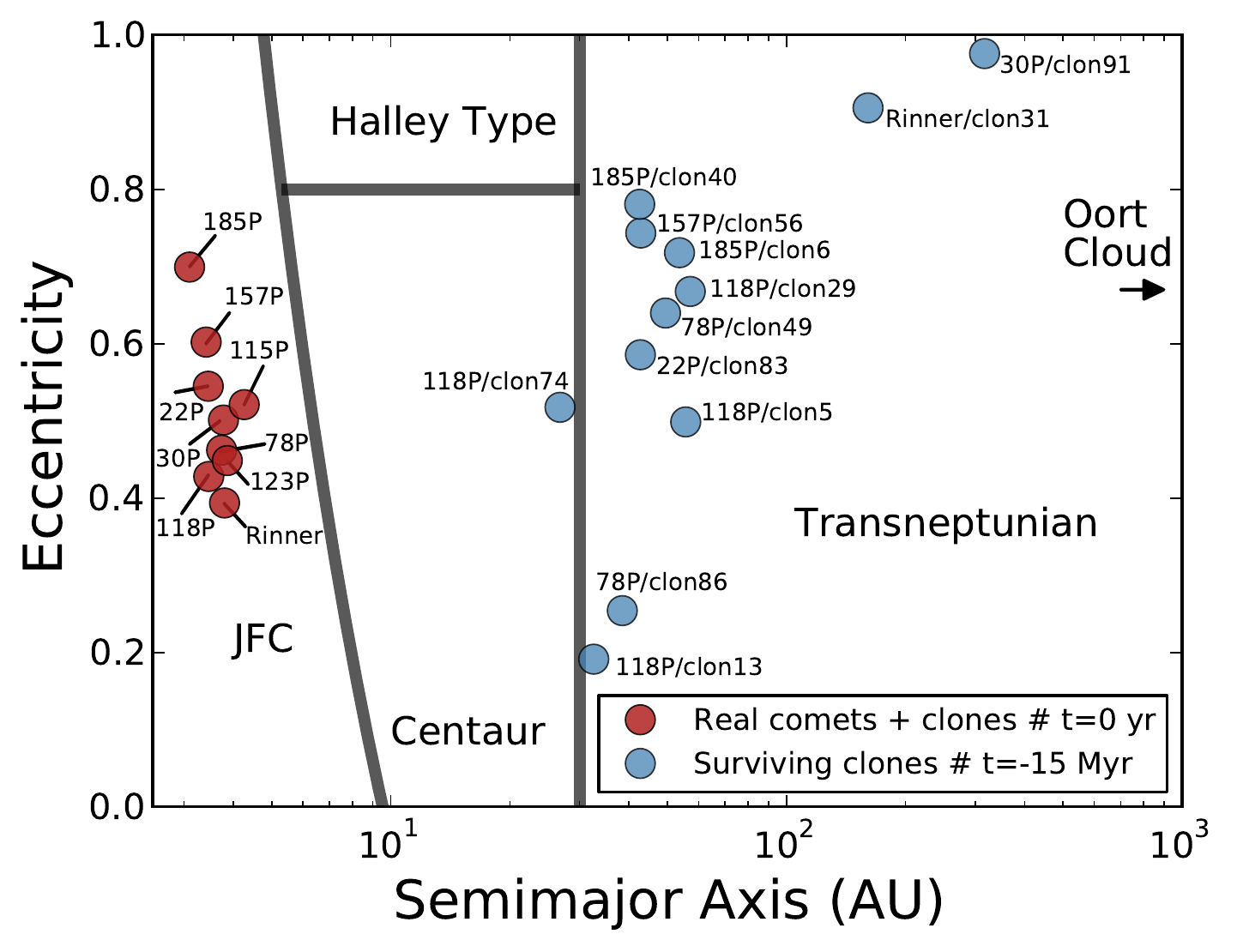}
          \caption{Time evolution of the 900 initial particles in the 15 Myr backward integration. Red circles are
                   the real comets and their clones (in the same location for t=0 yr, which is current position, just 2$\sigma$ dispersion in the orbital parameters), and the blue circles are
                   the surving particles after 15 Myr.                    
                    }
         \label{dinamica_total}
   \end{figure}

From the dynamical analysis, we determine that, just 12 
of the initial 900 particles (9 real comets + 99 clones per each one) survived after a 15 Myr backward integration, which means 1.3$\%$.
This result agrees with \citet{levison1994}, who concluded 
that just 11$\pm4$ particles, or $1.5\pm0.6\%$, remained in the Solar System after integration from their dynamical study. In Fig. \ref{dinamica_total}, we show 
the surviving clones in the $a$-$e$ plane, where just one of the clones is in the Centaur region (118P/clon74) and the rest of them are in the transneptunian region.
Two of the clones have $a>100$ AU with a very high eccentricity ($e>0.9$), 30P/clon91 and Rinner/clon31. On the other hand, 
there are two comets with low eccentricity, 78P/clon86 and 118P/clon13, with $e<0.25$. The rest of the surviving clones have intermediate values of eccentricity, 
$0.45<e<0.8$, and it seems that these comets are in a transition state between Kuiper Belt and the Scattered Disk objects \citep{levison1997}. 

\begin{figure}
      \includegraphics[width=1\columnwidth]{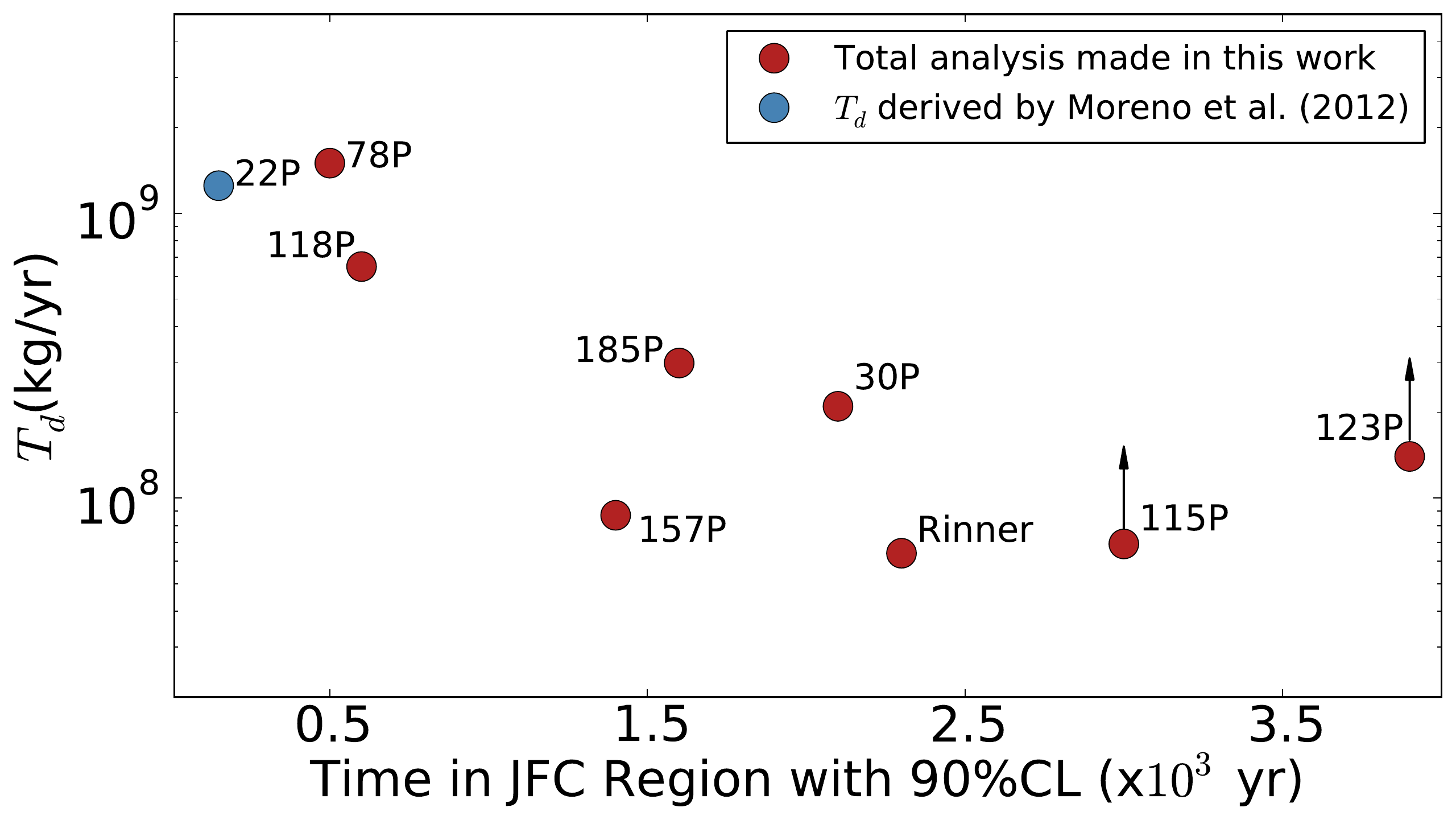}
          \caption{Annual dust production rate of our targets obtained in the dust analysis (see section 4) versus 
                   the time in the JFCs region with a 
                   $90\%$CL derived in dynamical studies (see section 5). The comets with arrows mean the 
                   $T_{d}$ given for them are lower limits.
                    }
         \label{final}
   \end{figure}

In addition, we derived the time spent by all the comets under study in the JFCs region with a 90$\%$ of confidence level. We can see that the youngest is 22P, 
followed by 78P, and 118P ($\sim100$,$\sim500$, and $\sim600$ yr respectively). On the other hand, the oldest comet in our sample is 123P with $\sim3.9\times10^{3}$ yr.
This result is shown in Fig. \ref{final}, where we relate the annual dust production rate ($T_{d}$, see section 4) within the time spent in the JFCs region for each comet.
It
seems that the most active comets in our sample are at the same time the youngest ones, which are, 22P, 78P and 118P.  

\section{Summary and conclusions}

            We presented optical observations, which were carried out at Sierra Nevada Observatory on 1.52 m telescope, of eight JFCs comets during their last perihelion passage: 
            30P/Reinmuth 1, 78P/Gehrels 2, 115P/Maury, 118P/Shoemaker-Levy 4, 123P/West-Hartley, 157P/Tritton, 185/Petriew, and 
            P/2011 W2 (Rinner). We also benefited from $Af\rho$ curves of these targets along $\sim{300}$ days around perihelion, 
            which is provided by \emph{Cometas-Obs}. We used our Monte Carlo dust tail code \citep[e.g.][]{moreno2009} to derive the dust properties 
            of our targets. These properties were dust loss rate, ejection velocities of particles, and size distribution
            of particles, where we gave the minimum and maximum size of particles and 
            the power index of the size distribution $\delta$. We also obtained the overall emission pattern for each comet, which could be either
            isotropic or anisotropic. When the ejection was derived as anisotropic, we could estimate the location of the active areas on the surface
            and the rotational parameters given by $\phi$ and $I$. From this analysis, we have determined three categories according 
            to the amount of dust emitted:
           
           \begin{enumerate}
            
            \item Weakly active: 115P, 157P, and Rinner with an annual production rate $T_{d}<1\times10^{8}$ kg yr$^{-1}$.
            \item Moderately active: 30P, 123P, and 185P with an annual production rate of $T_{d}=1-3\times10^{8}$ kg yr$^{-1}$.
            \item Highly active: 78P and 118P with values $T_{d}>8\times10^{8}$ kg yr$^{-1}$. In addition to these targets, we also considered for our purposes
                  the results of the dust characterization given in a previous work by \citet{moreno2012} for the comet 22P/Kopff. 
                  For this object, the annual production rate 
                  was derived as $T_{d}=1.24\times10^{9}$ kg yr$^{-1}$, which allowed us introduced it in this category.
            \end{enumerate}
            These results should be regarded as lower limits because largest size particles are not tightly constrained.\\

            The second part of our study was the determination of the dynamical evolution followed by the comets of the sample in the last 1 Myr. With this 
            purpose, we used the numerical integrator developed by \citet{chambers1999}. In that case, we neglected the non-gravitational forces
            due to the little contribution of a single jet in the motion of our targets. We derived its maximum influence over $a$ as 0.33 AU
            during the lifetime of the sublimation jet. To make a statistical study of the dynamical evolution, we used 99 clones
            with 2$\sigma$ dispersion in the orbital parameters ($a$, $e$, and $i$) and the real one. Thus, we had 100 test particles to 
            determine, which were the most visited regions by each comet and when. That analysis allowed us to determine how long
            these comets spent as members of JFCs, region of special interest because it is supposed that this is the place where
            the comets became active by sublimating the ices trapped in the nucleus, which belong to the primitive chemical components
            of the Solar System when was formed. From the dynamical study, we inferred  
            that our targets were relatively young in the JFC region with ages between $100<t<4000$ years, and all of them have a Centaur and
            Transneptunian past, as expected.\\
            
            The last point in our conclusions led us to relate the results in the previous points. In Fig. \ref{final}, we plotted 
            each comet by attending to the averaged dust production rate [kg/yr] derived in the dust characterizations (see table \ref{DPII} in section 4) 
            and the time spent in the JFCs region that are obtained 
            in the dynamical analysis (see section 5). Attending to that figure, 
            we concluded that the most active comets in our target list are at the same time 
            the youngest ones (22P, 78P, and 118P). Although the other targets showed a similar trend in general, there were some exceptions (e.g., 157P and 123P) that
            prevent us from reaching a firm conclusion. A more extended study of this kind would then be desirable.
            

\begin{acknowledgements}
      We thank to F. Aceituno, V. Casanova, and A. Sota for their support as 
      staff members in the Sierra Nevada Observatory. The amateur astronomical association \emph{Cometas-Obs}
      and the full grid of observers who spend the nights looking for comets. Also we want to 
      thank Dr. Chambers for his help using his numerical integrator, and the anonymous referee for comments and suggestion for improving the paper.
      This work was supported by contracts \emph{AYA2012-3961-CO2-01} and \emph{FQM-4555} (Proyecto de Excelencia, Junta de Andalucia).
\end{acknowledgements}


\bibliographystyle{aa}
\bibliography{curroreferencias}

\begin{thebibliography}{36}
\expandafter\ifx\csname natexlab\endcsname\relax\def\natexlab#1{#1}\fi

\bibitem[{{A'Hearn} {et~al.}(2005){A'Hearn}, {Belton}, {Delamere}, {Kissel},
  {Klaasen}, {McFadden}, {Meech}, {Melosh}, {Schultz}, {Sunshine}, {Thomas},
  {Veverka}, {Yeomans}, {Baca}, {Busko}, {Crockett}, {Collins}, {Desnoyer},
  {Eberhardy}, {Ernst}, {Farnham}, {Feaga}, {Groussin}, {Hampton}, {Ipatov},
  {Li}, {Lindler}, {Lisse}, {Mastrodemos}, {Owen}, {Richardson}, {Wellnitz}, \&
  {White}}]{ahearn2005}
{A'Hearn}, M.~F., {Belton}, M.~J.~S., {Delamere}, W.~A., {et~al.} 2005,
  Science, 310, 258

\bibitem[{{A'Hearn} {et~al.}(1984){A'Hearn}, {Schleicher}, {Millis}, {Feldman},
  \& {Thompson}}]{ahearn1984}
{A'Hearn}, M.~F., {Schleicher}, D.~G., {Millis}, R.~L., {Feldman}, P.~D., \&
  {Thompson}, D.~T. 1984, \aj, 89, 579

\bibitem[{{Brownlee} {et~al.}(2004){Brownlee}, {Horz}, {Newburn}, {Zolensky},
  {Duxbury}, {Sandford}, {Sekanina}, {Tsou}, {Hanner}, {Clark}, {Green}, \&
  {Kissel}}]{brownlee2004}
{Brownlee}, D.~E., {Horz}, F., {Newburn}, R.~L., {et~al.} 2004, Science, 304,
  1764

\bibitem[{{Burns} {et~al.}(1979){Burns}, {Lamy}, \& {Soter}}]{burns1979}
{Burns}, J.~A., {Lamy}, P.~L., \& {Soter}, S. 1979, \icarus, 40, 1

\bibitem[{{Carry}(2012)}]{carry2012}
{Carry}, B. 2012, \planss, 73, 98

\bibitem[{{Chambers}(1999)}]{chambers1999}
{Chambers}, J.~E. 1999, \mnras, 304, 793

\bibitem[{{Davidsson} \& {Gutierrez}(2004)}]{davidsson2004}
{Davidsson}, B.~J.~R. \& {Gutierrez}, P.~J. 2004, in Bulletin of the American
  Astronomical Society, Vol.~36, AAS/Division for Planetary Sciences Meeting
  Abstracts \#36, 1118

\bibitem[{{Edoh}(1983)}]{edoh1983}
{Edoh}, O. 1983, Univ. Arizona

\bibitem[{{Finson} \& {Probstein}(1968)}]{finson&probstein1968a}
{Finson}, M.~J. \& {Probstein}, R.~F. 1968, \apj, 154, 327

\bibitem[{{Fulle} {et~al.}(2010){Fulle}, {Colangeli}, {Agarwal}, {Aronica},
  {Della Corte}, {Esposito}, {Gr{\"u}n}, {Ishiguro}, {Ligustri}, {Lopez
  Moreno}, {Mazzotta Epifani}, {Milani}, {Moreno}, {Palumbo}, {Rodr{\'i}guez
  G{\'o}mez}, \& {Rotundi}}]{fulle2010}
{Fulle}, M., {Colangeli}, L., {Agarwal}, J., {et~al.} 2010, \aap, 522, A63

\bibitem[{{Grun} {et~al.}(1985){Grun}, {Zook}, {Fechtig}, \&
  {Giese}}]{grun1985}
{Grun}, E., {Zook}, H.~A., {Fechtig}, H., \& {Giese}, R.~H. 1985, \icarus, 62,
  244

\bibitem[{{Hartogh} {et~al.}(2011){Hartogh}, {Lis}, {Bockel{\'e}e-Morvan}, {de
  Val-Borro}, {Biver}, {K{\"u}ppers}, {Emprechtinger}, {Bergin}, {Crovisier},
  {Rengel}, {Moreno}, {Szutowicz}, \& {Blake}}]{hartogh2011}
{Hartogh}, P., {Lis}, D.~C., {Bockel{\'e}e-Morvan}, D., {et~al.} 2011, \nat,
  478, 218

\bibitem[{{Hsieh} {et~al.}(2012{\natexlab{a}}){Hsieh}, {Yang}, \&
  {Haghighipour}}]{hsieh2012a}
{Hsieh}, H.~H., {Yang}, B., \& {Haghighipour}, N. 2012{\natexlab{a}}, \apj,
  744, 9

\bibitem[{{Hsieh} {et~al.}(2012{\natexlab{b}}){Hsieh}, {Yang}, {Haghighipour},
  {Novakovi{\'c}}, {Jedicke}, {Wainscoat}, {Denneau}, {Abe}, {Chen},
  {Fitzsimmons}, {Granvik}, {Grav}, {Ip}, {Kaluna}, {Kinoshita}, {Kleyna},
  {Knight}, {Lacerda}, {Lisse}, {Maclennan}, {Meech}, {Micheli}, {Milani},
  {Pittichov{\'a}}, {Schunova}, {Tholen}, {Wasserman}, {Burgett}, {Chambers},
  {Heasley}, {Kaiser}, {Magnier}, {Morgan}, {Price}, {J{\o}rgensen}, {Dominik},
  {Hinse}, {Sahu}, \& {Snodgrass}}]{hsieh2012b}
{Hsieh}, H.~H., {Yang}, B., {Haghighipour}, N., {et~al.} 2012{\natexlab{b}},
  \aj, 143, 104

\bibitem[{{Jewitt}(2009)}]{jewitt2009}
{Jewitt}, D. 2009, \aj, 137, 4296

\bibitem[{{Jockers}(1997)}]{jockers1997}
{Jockers}, K. 1997, Earth Moon and Planets, 79, 221

\bibitem[{{Keller} {et~al.}(1986){Keller}, {Arpigny}, {Barbieri}, {Bonnet},
  {Cazes}, {Coradini}, {Cosmovici}, {Delamere}, {Huebner}, {Hughes}, {Jamar},
  {Malaise}, {Reitsema}, {Schmidt}, {Schmidt}, {Seige}, {Whipple}, \&
  {Wilhelm}}]{keller1986}
{Keller}, H.~U., {Arpigny}, C., {Barbieri}, C., {et~al.} 1986, \nat, 321, 320

\bibitem[{{Kolokolova} {et~al.}(2004){Kolokolova}, {Hanner},
  {Levasseur-Regourd}, \& {Gustafson}}]{kolokolova2004}
{Kolokolova}, L., {Hanner}, M.~S., {Levasseur-Regourd}, A.-C., \& {Gustafson},
  B.~{\AA}.~S. 2004, {Physical properties of cometary dust from light
  scattering and thermal emission}, ed. G.~W. {Kronk}, 577--604

\bibitem[{{Krolikowska} {et~al.}(1998){Krolikowska}, {Sitarski}, \&
  {Szutowicz}}]{krolikowska1998}
{Krolikowska}, M., {Sitarski}, G., \& {Szutowicz}, S. 1998, \actaa, 48, 91

\bibitem[{{Lacerda}(2013)}]{lacerda2013}
{Lacerda}, P. 2013, \mnras, 428, 1818

\bibitem[{{Lamy} {et~al.}(2004){Lamy}, {Toth}, {Fernandez}, \&
  {Weaver}}]{lamy2004}
{Lamy}, P.~L., {Toth}, I., {Fernandez}, Y.~R., \& {Weaver}, H.~A. 2004, {The
  sizes, shapes, albedos, and colors of cometary nuclei}, ed. G.~W. {Kronk},
  223--264

\bibitem[{{Levison} \& {Duncan}(1994)}]{levison1994}
{Levison}, H.~F. \& {Duncan}, M.~J. 1994, \icarus, 108, 18

\bibitem[{{Levison} \& {Duncan}(1997)}]{levison1997}
{Levison}, H.~F. \& {Duncan}, M.~J. 1997, \icarus, 127, 13

\bibitem[{{Lowry} \& {Weissman}(2003)}]{lowry2003}
{Lowry}, S.~C. \& {Weissman}, P.~R. 2003, \icarus, 164, 492

\bibitem[{{Mazzotta Epifani} \& {Palumbo}(2011)}]{mazzotta2011}
{Mazzotta Epifani}, E. \& {Palumbo}, P. 2011, \aap, 525, A62

\bibitem[{{Meech} \& {Jewitt}(1987)}]{meech1987}
{Meech}, K.~J. \& {Jewitt}, D.~C. 1987, \aap, 187, 585

\bibitem[{{Monet} {et~al.}(2003){Monet}, {Levine}, {Canzian}, {Ables}, {Bird},
  {Dahn}, {Guetter}, {Harris}, {Henden}, {Leggett}, {Levison}, {Luginbuhl},
  {Martini}, {Monet}, {Munn}, {Pier}, {Rhodes}, {Riepe}, {Sell}, {Stone},
  {Vrba}, {Walker}, {Westerhout}, {Brucato}, {Reid}, {Schoening}, {Hartley},
  {Read}, \& {Tritton}}]{monet2003}
{Monet}, D.~G., {Levine}, S.~E., {Canzian}, B., {et~al.} 2003, \aj, 125, 984

\bibitem[{{Moreno}(2009)}]{moreno2009}
{Moreno}, F. 2009, \apjs, 183, 33

\bibitem[{{Moreno} {et~al.}(2011){Moreno}, {Lara}, {Licandro}, {Ortiz}, {de
  Le{\'o}n}, {Al{\'i}-Lagoa}, {Ag{\'i}s-Gonz{\'a}lez}, \&
  {Molina}}]{moreno2011}
{Moreno}, F., {Lara}, L.~M., {Licandro}, J., {et~al.} 2011, \apjl, 738, L16

\bibitem[{{Moreno} {et~al.}(2012){Moreno}, {Pozuelos}, {Aceituno}, {Casanova},
  {Sota}, {Castellano}, \& {Reina}}]{moreno2012}
{Moreno}, F., {Pozuelos}, F., {Aceituno}, F., {et~al.} 2012, \apj, 752, 136

\bibitem[{{Schwehm} \& {Schulz}(1998)}]{schwehm1998}
{Schwehm}, G. \& {Schulz}, R. 1998, in Astrophysics and Space Science Library,
  Vol. 236, Laboratory astrophysics and space research, ed. P.~{Ehrenfreund},
  C.~{Krafft}, H.~{Kochan}, \& V.~{Pirronello}, 537

\bibitem[{{Scotti}(1994)}]{scotti1994}
{Scotti}, J.~V. 1994, in Bulletin of the American Astronomical Society,
  Vol.~26, American Astronomical Society Meeting Abstracts, 1375

\bibitem[{{Sekanina}(1981)}]{sekanina1981}
{Sekanina}, Z. 1981, Annual Review of Earth and Planetary Sciences, 9, 113

\bibitem[{{Soderblom} {et~al.}(2002){Soderblom}, {Becker}, {Bennett}, {Boice},
  {Britt}, {Brown}, {Buratti}, {Isbell}, {Giese}, {Hare}, {Hicks},
  {Howington-Kraus}, {Kirk}, {Lee}, {Nelson}, {Oberst}, {Owen}, {Rayman},
  {Sandel}, {Stern}, {Thomas}, \& {Yelle}}]{soderblom2002}
{Soderblom}, L.~A., {Becker}, T.~L., {Bennett}, G., {et~al.} 2002, Science,
  296, 1087

\bibitem[{{Sykes} {et~al.}(2004){Sykes}, {Gr{\"u}n}, {Reach}, \&
  {Jenniskens}}]{sykes2004}
{Sykes}, M.~V., {Gr{\"u}n}, E., {Reach}, W.~T., \& {Jenniskens}, P. 2004, {The
  interplanetary dust complex and comets}, ed. M.~C. {Festou}, H.~U. {Keller},
  \& H.~A. {Weaver}, 677--693

\bibitem[{{Tancredi} {et~al.}(2006){Tancredi}, {Fern{\'a}ndez}, {Rickman}, \&
  {Licandro}}]{tancredi2006}
{Tancredi}, G., {Fern{\'a}ndez}, J.~A., {Rickman}, H., \& {Licandro}, J. 2006,
  \icarus, 182, 527

\end{thebibliography}

\Online
\section{Online Material}

\appendix

\section{Orbital parameters of the comets}
In table \ref{table:or}, we show the orbital elements of the comets used during the dynamical studies in section 5. They are extracted from 
        JPL Horizons on-line Solar System data.

\begin{table*}
\caption{Orbital parameters of the short period comets under study.}             
\label{table:or}      
\centering          
\begin{tabular}{c c c c c c c}     
\hline\hline        
\noalign{\smallskip}                     
\multirow{2}{*}{Comet} & \multirow{2}{*}{e$\pm\sigma$} & a$\pm\sigma$ & i$\pm\sigma$ & node & peri & M \\
&  & (AU) & ($\degree$) & ($\degree$) & ($\degree$) & ($\degree$)  \\  
\hline  
\noalign{\smallskip} 
22P & 0.54493307 & 3.4557183 & 4.727895 & \multirow{2}{*}{120.86178} & \multirow{2}{*}{162.64134} & \multirow{2}{*}{206.26869} \\ \
JPL K154/2 & $\pm$9e-8 &  $\pm$4e-7 & $\pm$5e-6 &  & & \\ \hline
\noalign{\smallskip}
30P & 0.5011951 & 3.7754076 & 8.12265 & \multirow{2}{*}{119.74115} & \multirow{2}{*}{13.17407} & \multirow{2}{*}{61.00371} \\ \
JPL K103/1 & $\pm$2e-7 &  $\pm$2e-7 & $\pm$1e-5 &  & & \\ \hline
\noalign{\smallskip}
78P & 0.46219966 & 3.73541262 & 6.25491 & \multirow{2}{*}{210.55664} & \multirow{2}{*}{192.74376} & \multirow{2}{*}{76.36107} \\ \
JPL K114/7 & $\pm$9e-8 & $\pm$9e-8 & $\pm$1e-5 &  &  &   \\ \hline
\noalign{\smallskip}
115P & 0.5211645 & 4.2597076 & 11.687384 & \multirow{2}{*}{176.50309} & \multirow{2}{*}{120.41045} & \multirow{2}{*}{58.59101} \\ \
JPL 16 & $\pm$1e-7 & $\pm$5e-7 & $\pm$8e-6 &  &  &   \\ \hline
\noalign{\smallskip}
118P & 0.42817557 & 3.4654879 & 8.508415 & \multirow{2}{*}{151.77018} & \multirow{2}{*}{302.17416} & \multirow{2}{*}{143.78902} \\ \
JPL 49 & $\pm$8e-8 & $\pm$2e-7 & $\pm$5e-6 &  &  &   \\ \hline
\noalign{\smallskip}
123P & 0.4486260 & 3.8607445 & 15.35692 & \multirow{2}{*}{46.59827} & \multirow{2}{*}{102.82020} & \multirow{2}{*}{43.33196} \\ \
JPL 63 & $\pm$1e-7 & $\pm$4e-7 & $\pm$1e-5 &  &  &   \\ \hline
\noalign{\smallskip}
157P & 0.60217 & 3.4134 & 7.28480 & \multirow{2}{*}{300.01451} & \multirow{2}{*}{148.84243} & \multirow{2}{*}{174.36079} \\ \
JPL 28 & $\pm$1e-5 & $\pm$1e-4 & $\pm$7e-5 &  &  &   \\ \hline
\noalign{\smallskip}
185P & 0.6993216 & 3.0996991 & 14.00701 & \multirow{2}{*}{214.09101} & \multirow{2}{*}{181.94033} & \multirow{2}{*}{62.38997} \\ \
JPL 43 & $\pm$1e-7 & $\pm$1e-7 & $\pm$1e-5 &  &  &   \\ \hline
\noalign{\smallskip}
Rinner & 0.39372 & 3.79871& 13.77393 & \multirow{2}{*}{232.01759} & \multirow{2}{*}{221.06138} & \multirow{2}{*}{8.62661} \\ \
JPL 19 & $\pm$1e-5 & $\pm$5e-5 & $\pm$8e-5 &  &  &   \\ \hline
\end{tabular}
\end{table*}

\section{Dust environment of the comets in the sample}
\begin{figure}
      \includegraphics[width=1\columnwidth]{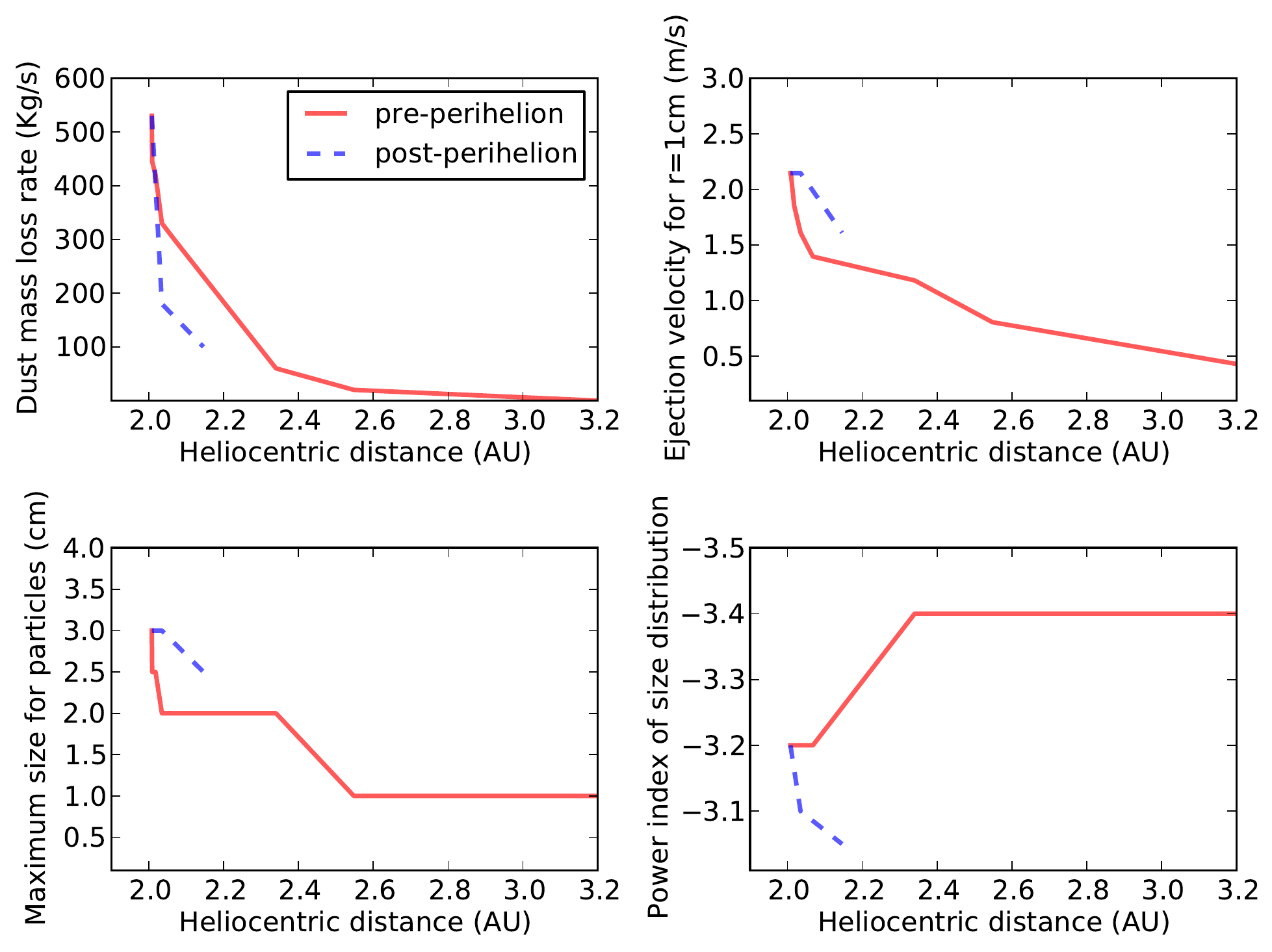}
          \caption{As in Fig. \ref{30P-DP}, but for comet 78P/Gehrels 2.
              }
         \label{78P-DP}
   \end{figure}

\begin{figure}
      \includegraphics[width=1\columnwidth]{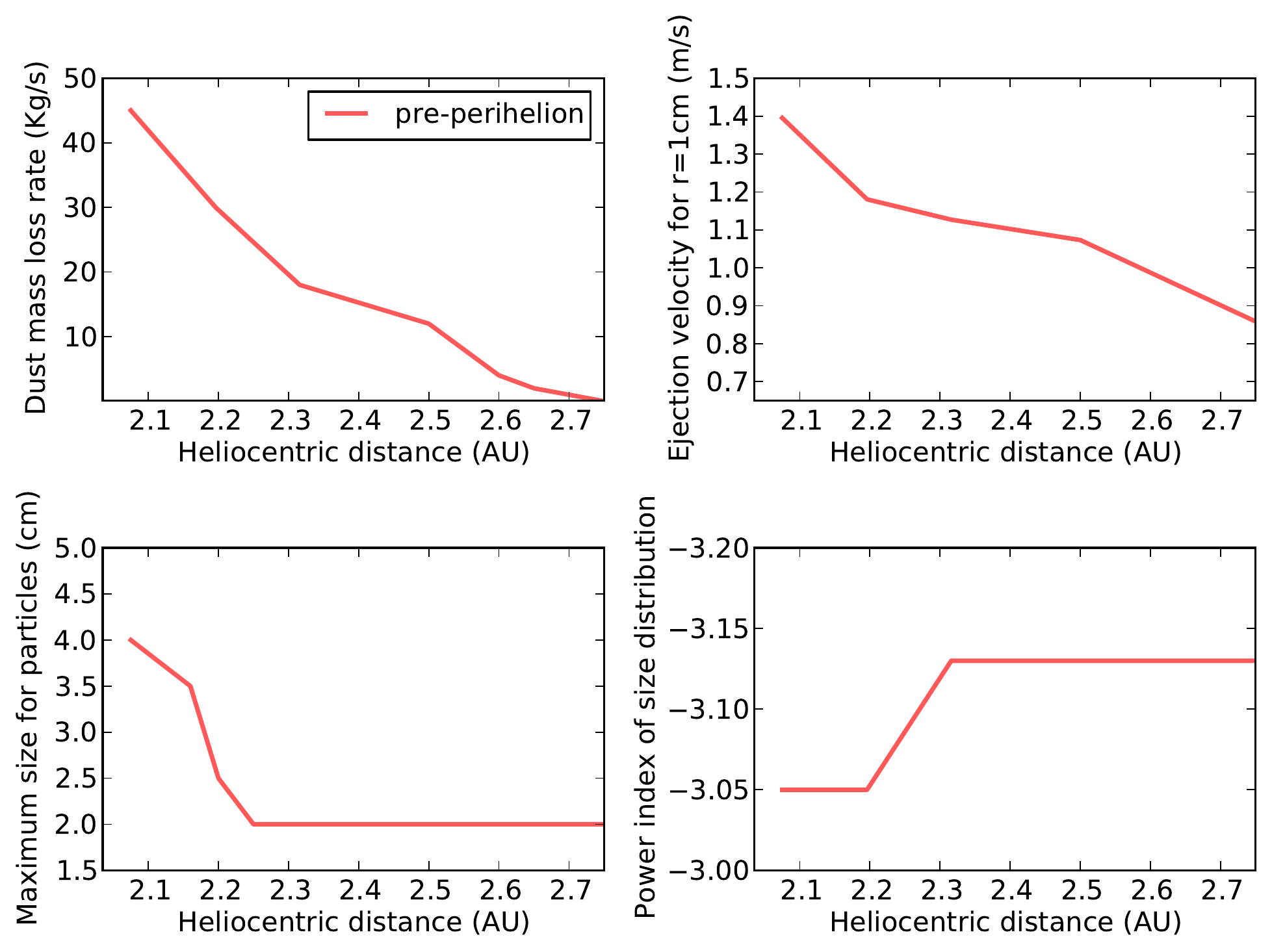}
          \caption{As in Fig. \ref{30P-DP}, but for comet 115P/Maury.
                    }
         \label{115P-DP}
   \end{figure}

\begin{figure}
      \includegraphics[width=1\columnwidth]{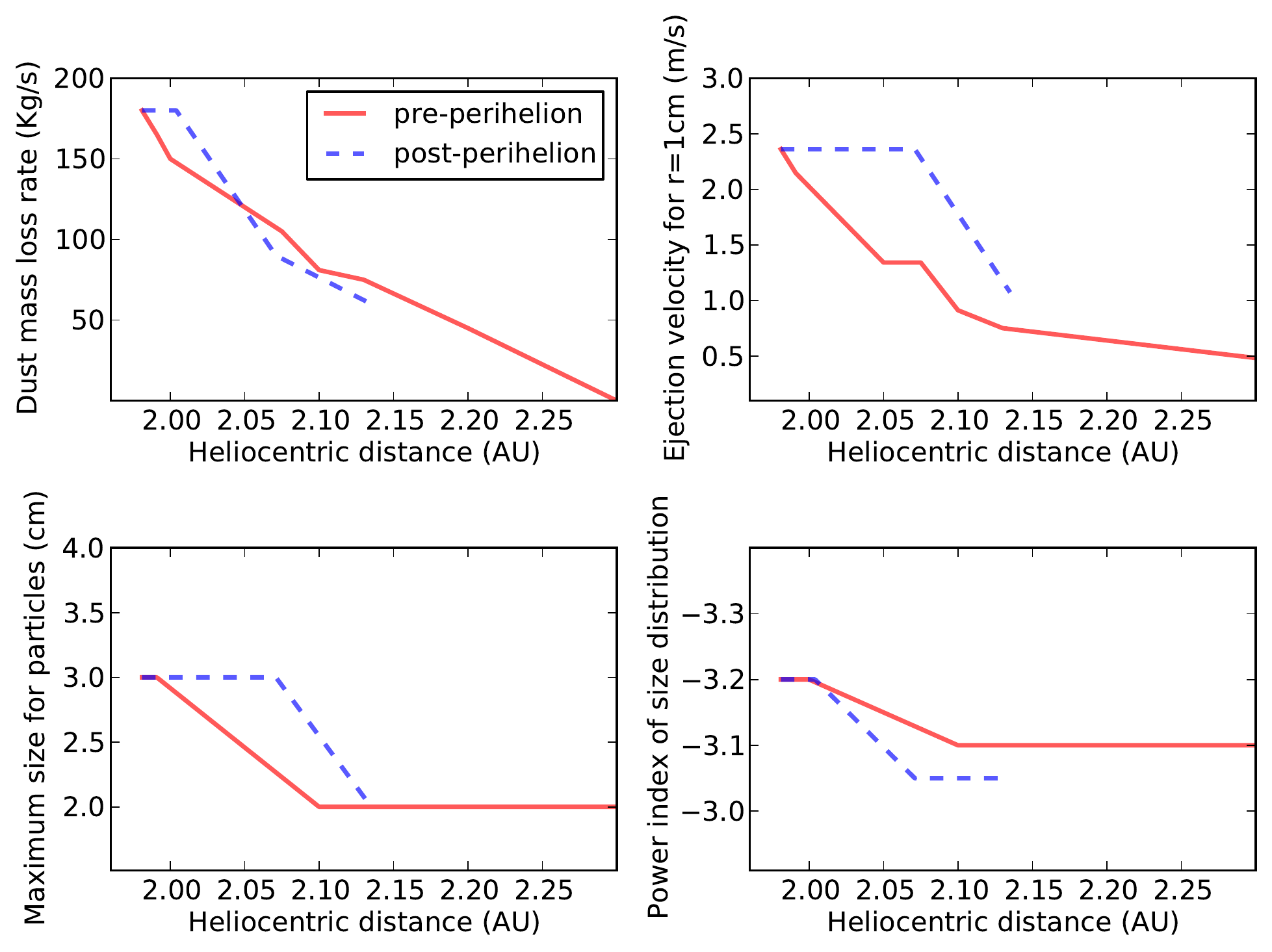}
          \caption{As in Fig. \ref{30P-DP}, but for comet 118P/Shoemaker-Levy 4.
                    }
         \label{118P-DP}
   \end{figure}

\begin{figure}
      \includegraphics[width=1\columnwidth]{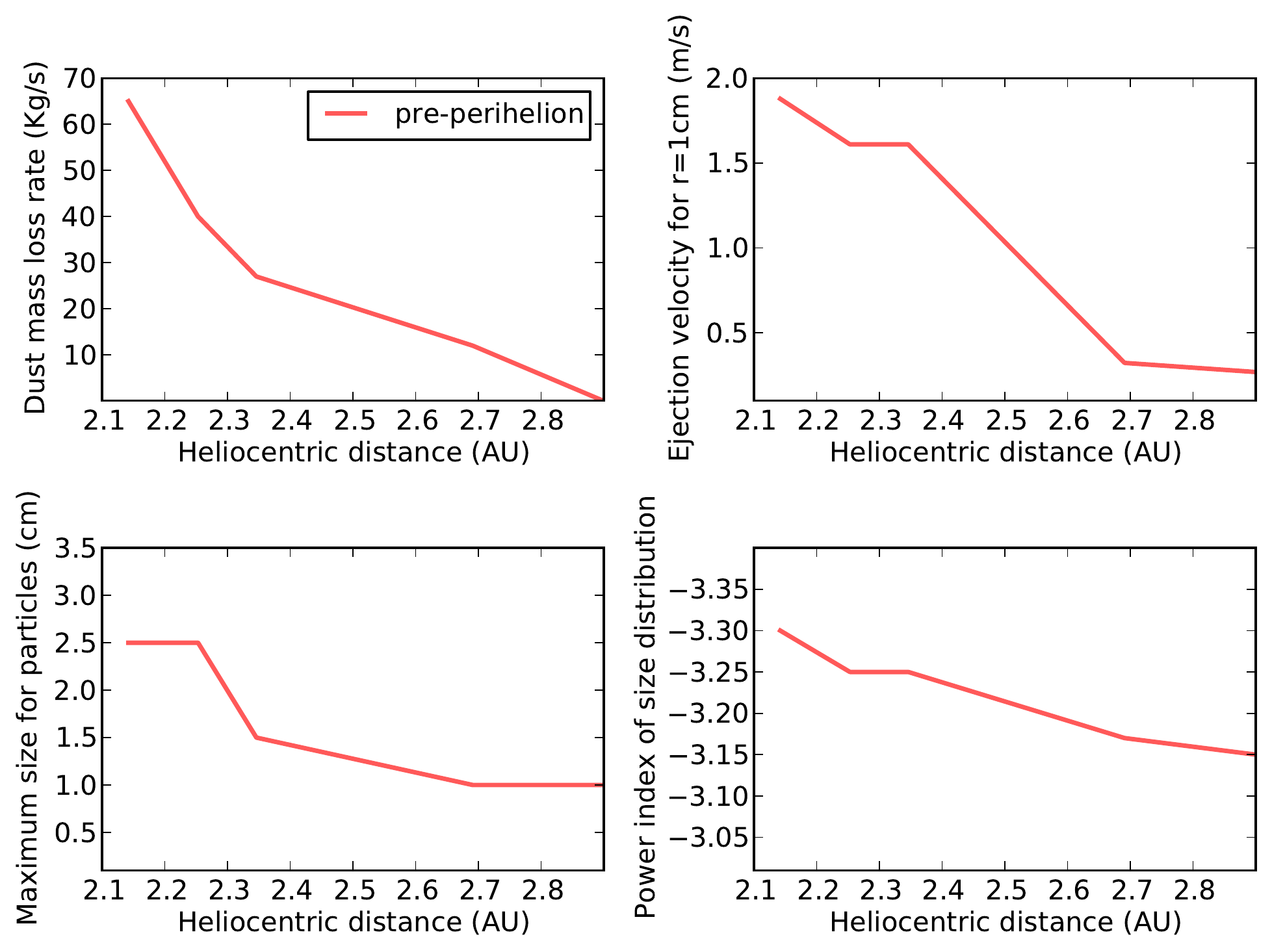}
          \caption{As in Fig. \ref{30P-DP}, but for comet 123P/West-Hartley.
                    }
         \label{123P-DP}
   \end{figure}

\begin{figure}
      \includegraphics[width=1\columnwidth]{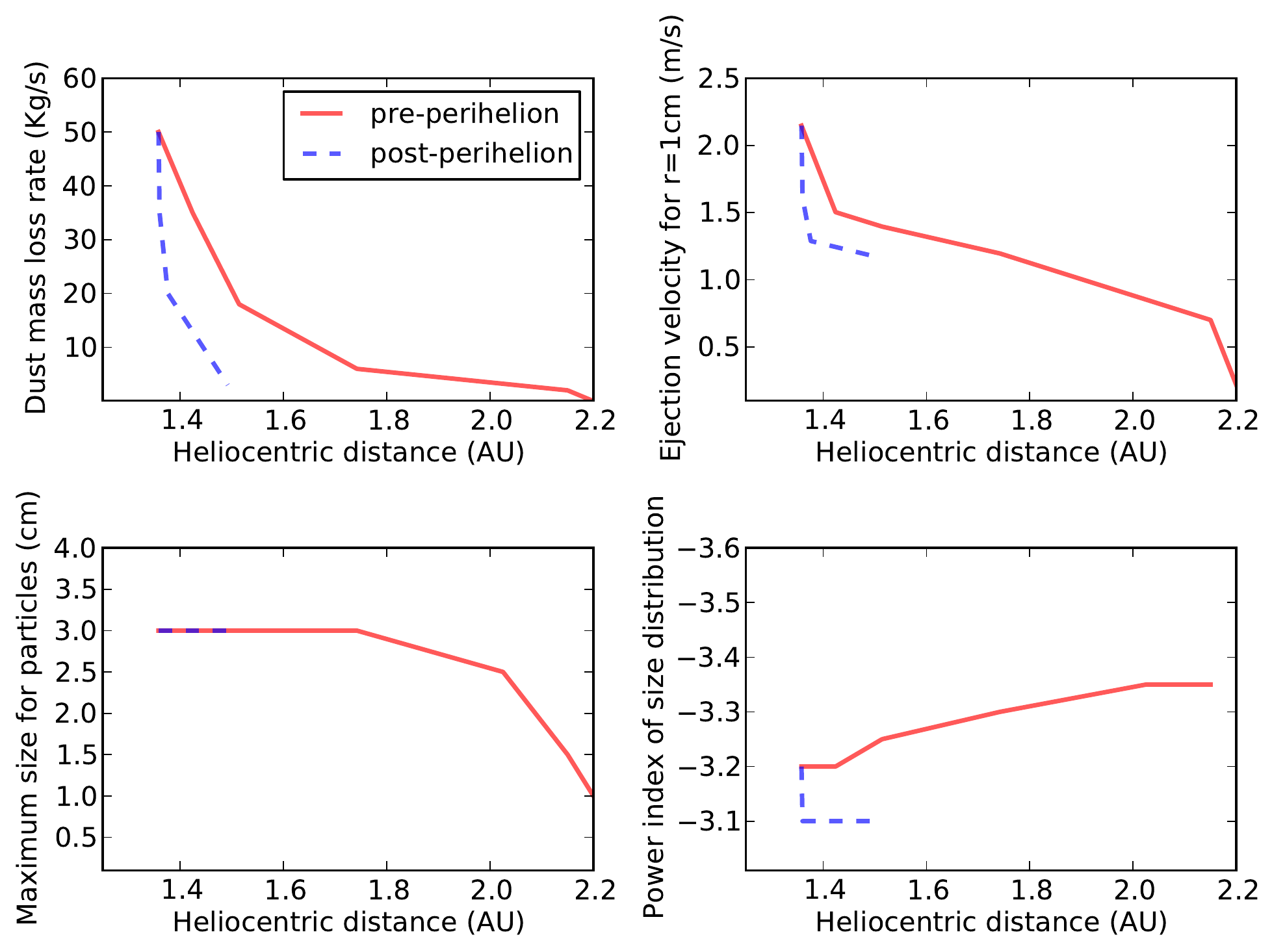}
          \caption{As in Fig. \ref{30P-DP}, but for comet 157P/Tritton.
                    }
         \label{157P-DP}
   \end{figure}

\begin{figure}
      \includegraphics[width=1\columnwidth]{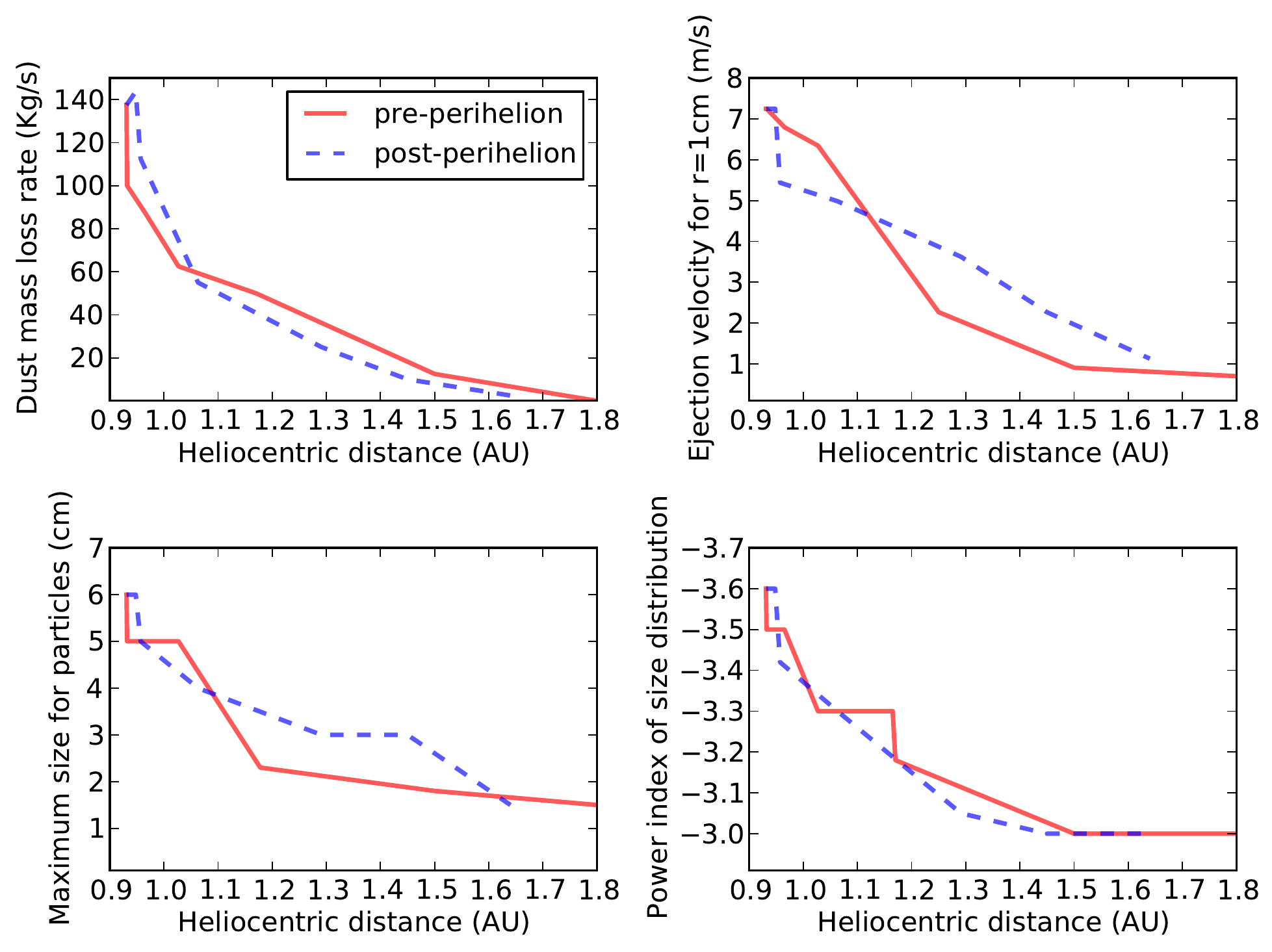}
          \caption{As in Fig. \ref{30P-DP}, but for comet 185P/Petriew.
                    }
         \label{185P-DP}
   \end{figure}

\begin{figure}
      \includegraphics[width=1\columnwidth]{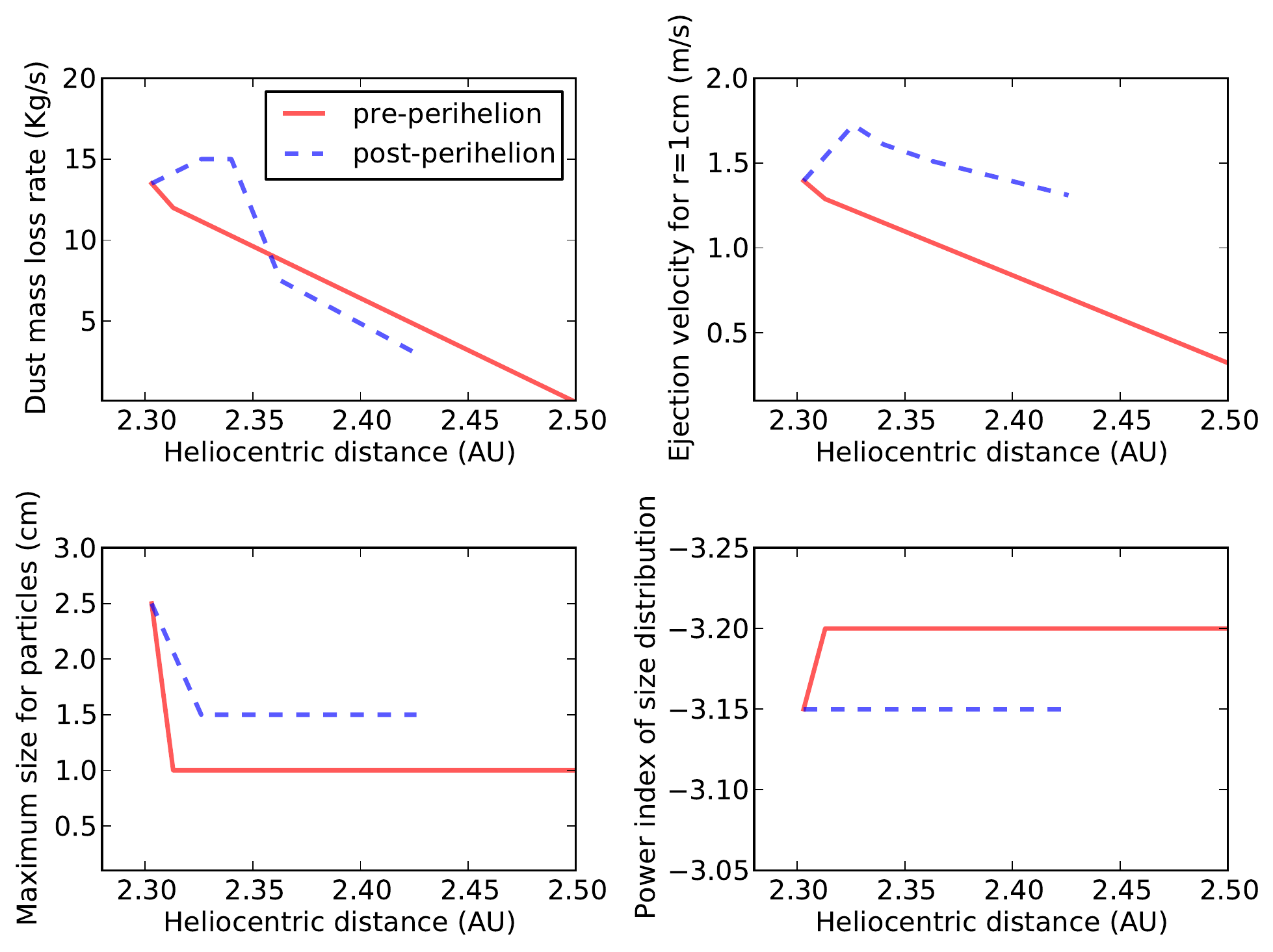}
          \caption{As in Fig. \ref{30P-DP}, but for comet P/2011 W2 (Rinner).
                    }
         \label{rinner-DP}
   \end{figure}

In this appendix, we present the evolution of the dust parameters versus heliocentric distance for each comet in the sample (figures \ref{78P-DP} to \ref{rinner-DP}).
These parameters are dust production rate [kg/s], ejection velocities for particles of r=1 cm glassy carbon spheres [m/s],
the maximum size of the particles [cm], and the power index of the size distribution ($\delta$). Solid red lines correspond to pre-perihelion, 
and dashed blue lines to post-perihelion.

\section{Comparison between observational data and models}
\begin{figure}
   \includegraphics[width=1\columnwidth]{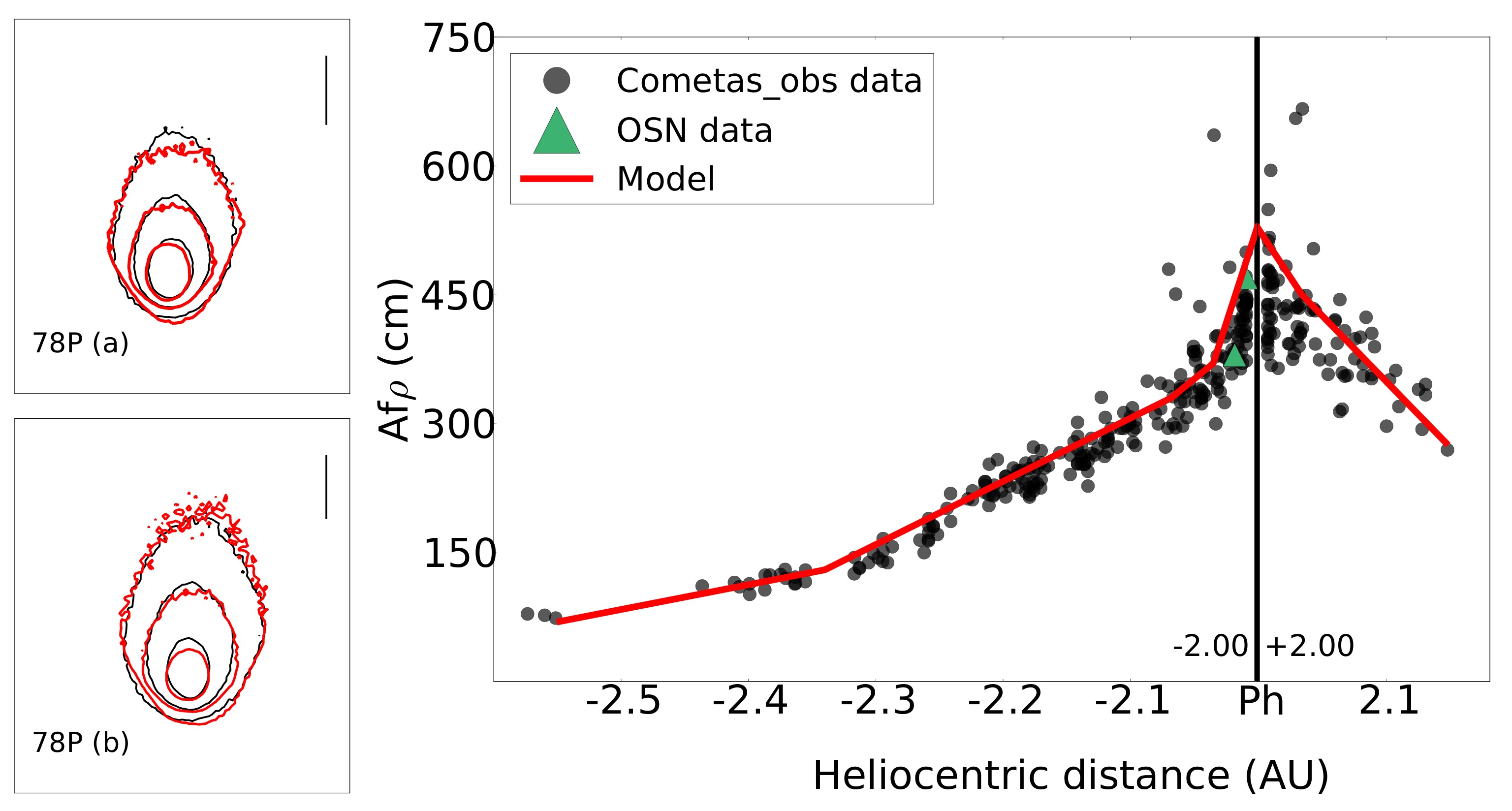}
      \caption{As in Fig. \ref{c30P}, but for comet 78P/Gehrels 2. Isophote fields:  (a) December 19, 2011. (b) January 4, 2012.
                In both cases the isophote levels are $0.55\times10^{-12}$, $2.65\times10^{-13}$, and 
                $1.35\times10^{-13}$ SDU.}
         \label{c78P}
   \end{figure}

\begin{figure}
   \includegraphics[width=1\columnwidth]{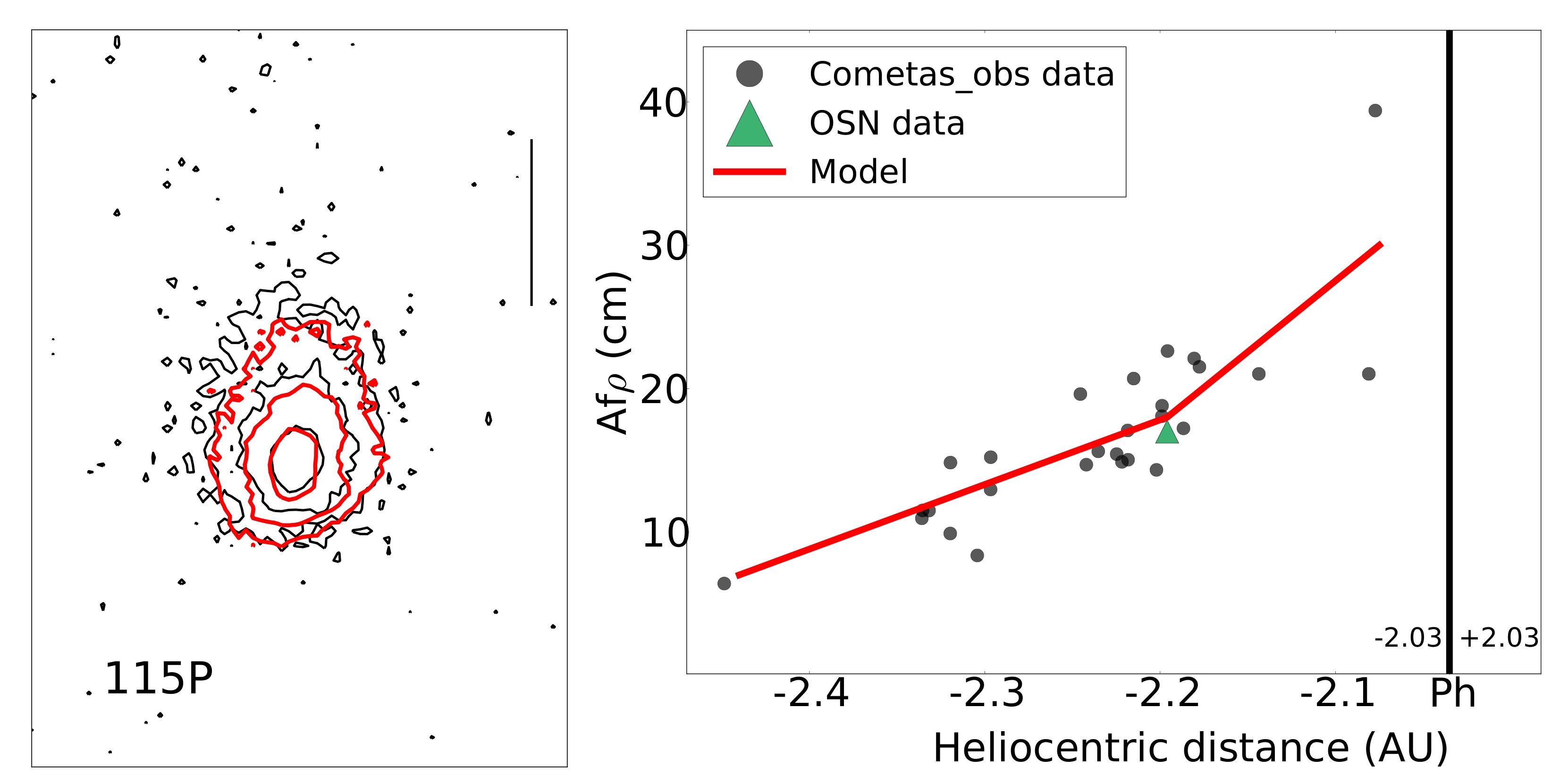}
      \caption{As in Fig. \ref{c30P}, but for comet 115P/Maury. Isophote fields: July 15, 2011. 
               Isophote levels are $1.00\times10^{-13}$, $3.00\times10^{-14}$, and 
               $1.30\times10^{-14}$ SDU.
              }
         \label{c115P}
   \end{figure}

\begin{figure}
   \includegraphics[width=1\columnwidth]{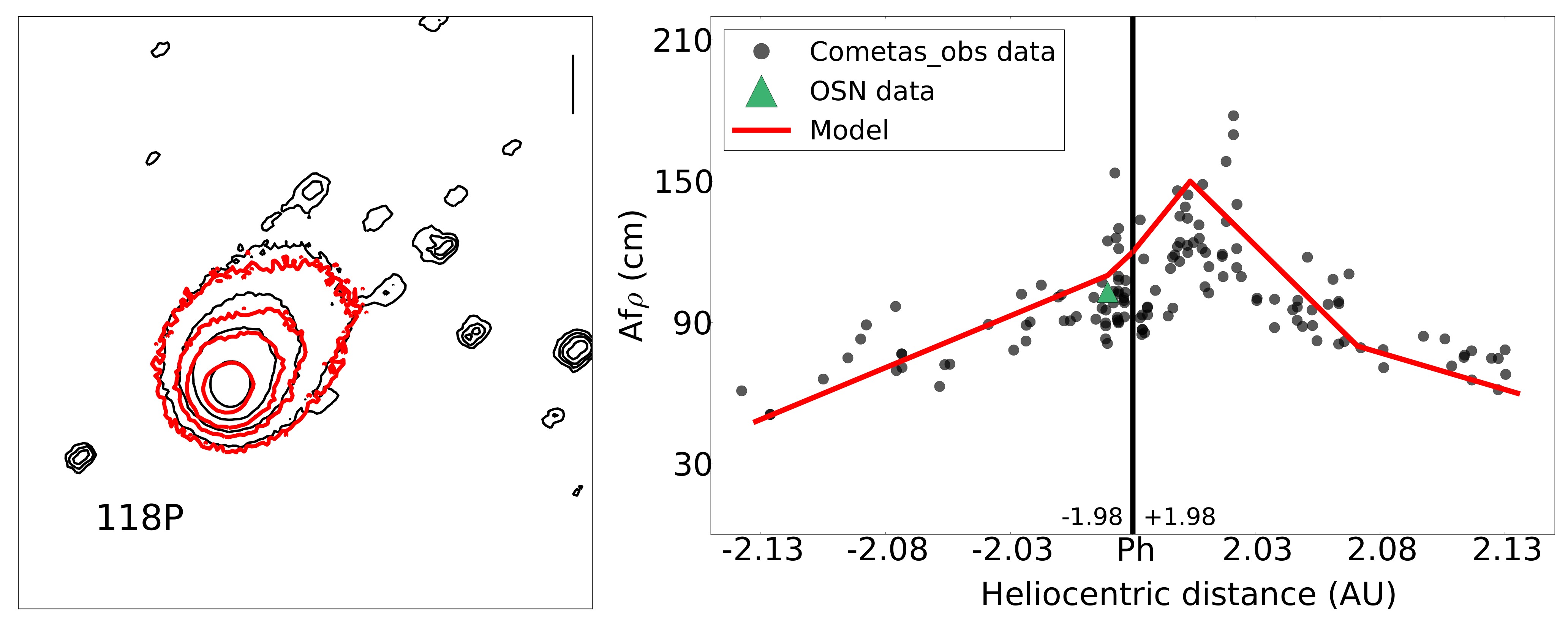}
      \caption{As in Fig. \ref{c30P}, but for comet 118P/Shoemaker-Levy 4. 
               Isophote fields: December 12, 2009. Isophote levels are $1.50\times10^{-13}$, $6.00\times10^{-14}$ 
              $3.50\times10^{-14}$, and $2.00\times10^{-14}$ SDU.
              }
         \label{c118P}
   \end{figure}

\begin{figure}
   \includegraphics[width=1\columnwidth]{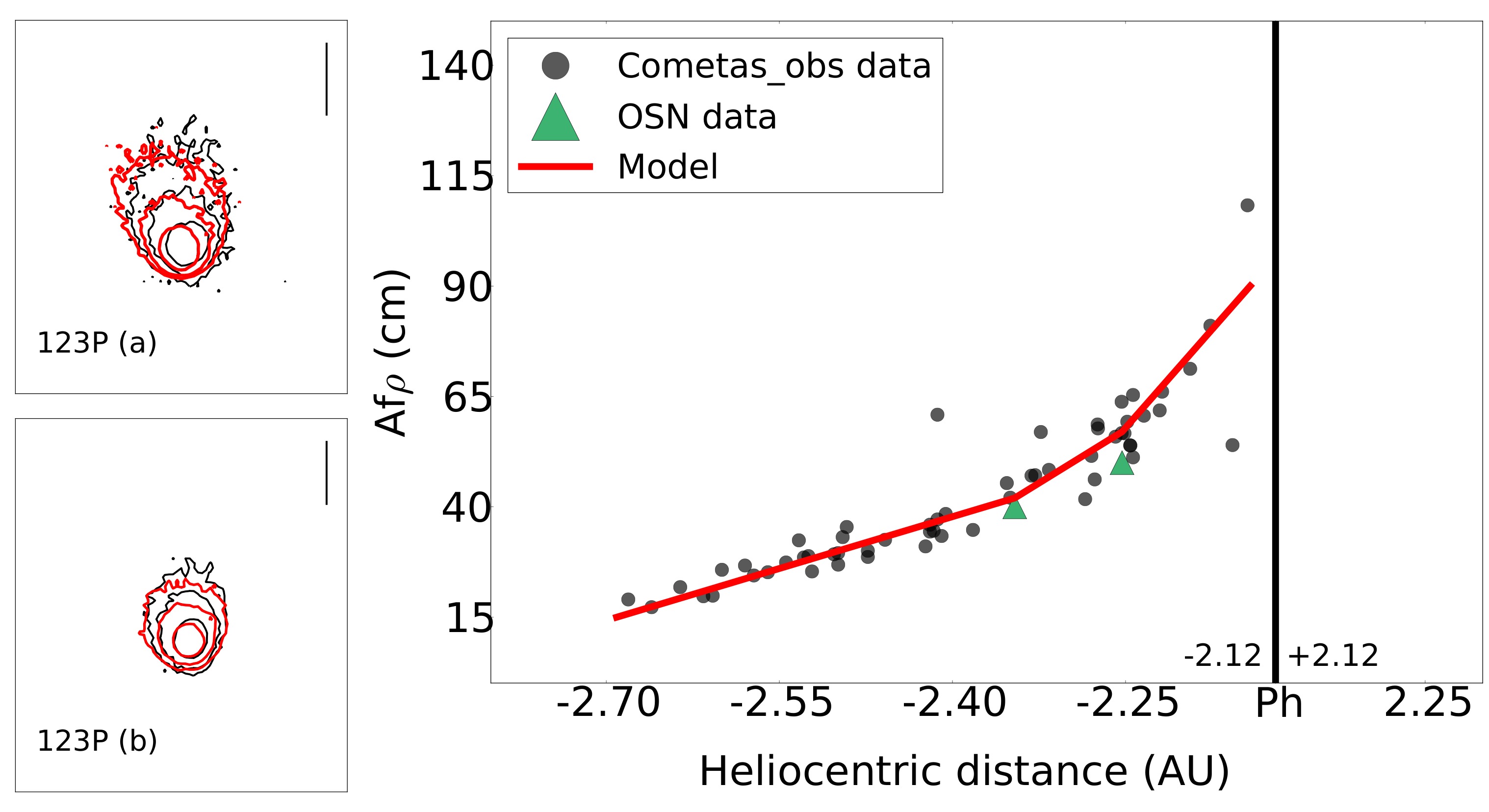}
      \caption{As in Fig. \ref{c30P}, but for comet 123P/West-Hartley. Isophote fields: (a) February 26, 2011. (b) March 31, 2011. 
              Isophote levels are 1.00$\times{10^{-13}}$, 0.35$\times{10^{-13}}$, and 0.15$\times{10^{-13}}$ SDU in (a) 
              and 1.50$\times{10^{-13}}$, 0.50$\times{10^{-13}}$, and 0.25$\times{10^{-13}}$ SDU in (b).
              }
         \label{c123P}
   \end{figure}

\begin{figure}
   \includegraphics[width=1\columnwidth]{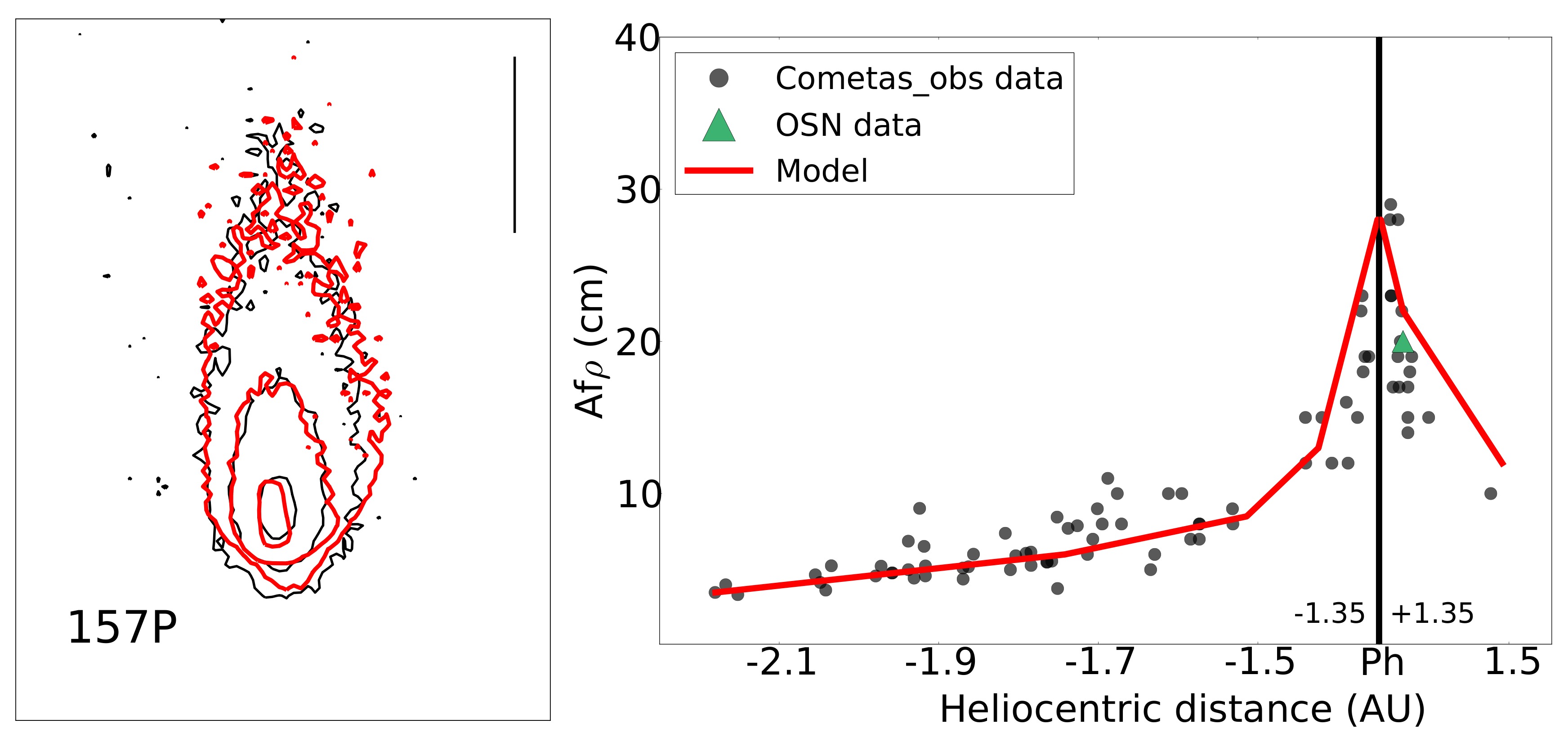}
      \caption{As in Fig. \ref{c30P}, but for comet 157P/Tritton. Isophote fields: March 10, 2010. Isophote levels are $6.00\times10^{-13}$, $0.75\times10^{-13}$, 
              and $2.65\times10^{-14}$ SDU.}
         \label{c157P}
   \end{figure}

\begin{figure}
   \includegraphics[width=1\columnwidth]{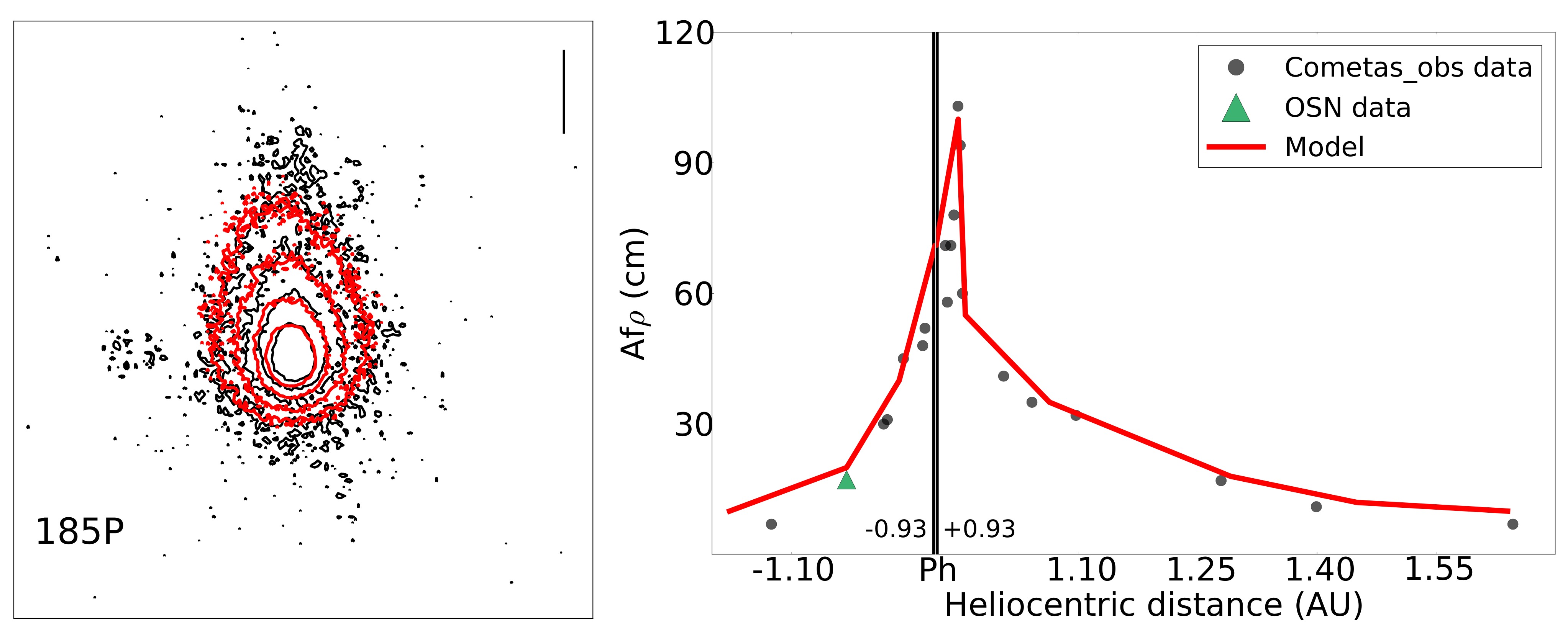}
      \caption{As in Fig. \ref{c30P}, but for comet 185P/Petriew. Isophote fields: July 15, 2012. Isophote levels are $1.80\times10^{-13}$, $1.00\times10^{-13}$ 
              $0.60\times10^{-13}$, and $0.35\times10^{-13}$ SDU.  
              }
         \label{c185P}
\end{figure}

\begin{figure}
   \includegraphics[width=1\columnwidth]{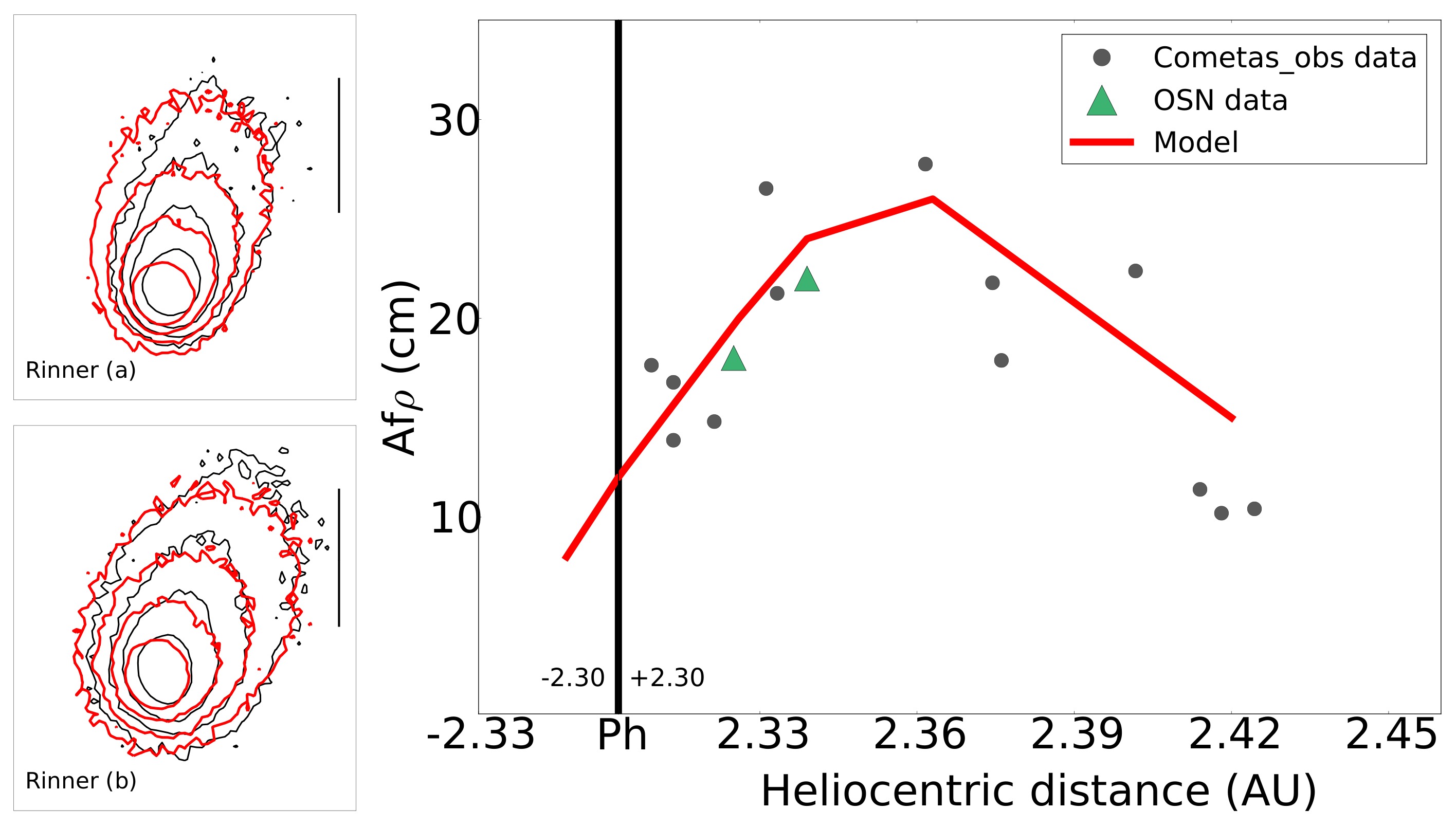}
      \caption{As in Fig. \ref{c30P}, but for comet P/2011 W2 (Rinner). Isophote fields: (a) December 22, 2011. 
              (b) January 4.
              In both cases the isophote are 6.00$\times{10^{-14}}$, 2.70$\times{10^{-14}}$,
              1.50$\times{10^{-14}}$, and 0.80$\times{10^{-14}}$ SDU. 
              }
         \label{crinner}
   \end{figure}

In this appendix (figures \ref{c78P} to \ref{crinner}), we show the comparison of the observational data and 
the models proposed in section 4, which describe the dust environments of the comets of the sample, as in Fig. \ref{c30P}. 

\end{document}